\documentclass[a4paper,fleqn,usenatbib]{mnras}

\usepackage{newtxtext,newtxmath}%
\usepackage[T1]{fontenc}%
\usepackage{ae,aecompl}%

\usepackage{color}
\usepackage{graphicx}
\usepackage{epsfig}

\usepackage{fancyhdr}
\usepackage{amsmath}
\usepackage{amssymb}
\usepackage{amsfonts}
\usepackage{typearea}
\usepackage{float}
\usepackage[bf]{caption}
\usepackage{here} 
\usepackage{subfig}
\usepackage{rotating}
\usepackage{hyperref}

\title[Morphology, Environment, and Satellites]{The Morphology-Density-Relation: Impact on the Satellite Fraction}

\author[A. F. Teklu et al.]{Adelheid F. Teklu$^{1,2}$\thanks{E-mail:
ateklu@usm.lmu.de}, Rhea-Silvia Remus$^{1}$, Klaus Dolag$^{1,3}$, \& Andreas Burkert$^{1,4}$\\
$^{1}$Universit\"ats-Sternwarte M\"unchen, Fakult\"at f\"ur Physik, LMU Munich, Scheinerstr. 1, D-81679 M\"unchen, Germany\\
$^{2}$Excellence Cluster Universe, Bolzmannstr. 2, D-85748 Garching, Germany\\
$^{3}$Max-Planck-Institute for Astrophysics, Karl-Schwarzschildstr. 1, D-85741 Garching, Germany\\
$^{4}$Max-Planck-Institute for Extraterrestrial Physics, Giessenbachstr. 1, D-85748 Garching, Germany}

\date{Accepted XXX. Received YYY; in original form ZZZ}

\pubyear{2017}

\begin{document}
\label{firstpage}

\pagerange{\pageref{firstpage}--\pageref{lastpage}} 

\maketitle

\begin{abstract}

In the past years several authors studied the abundance of satellites
around galaxies in order to better estimate, for example,
the halo masses of the host galaxies.
To investigate this connection, we analyze a sample of galaxies with
stellar masses $M_\mathrm{star} \geq 10^{10} M_{\odot}$, taken from the
hydrodynamical cosmological simulation suit Magneticum. 
We find that the satellite fraction of centrals is independent of their 
morphology. With the exception of very massive galaxies at low redshift, 
our results do not support the assumption that the dark matter haloes 
of spheroidal galaxies are significantly more massive than those of 
disc galaxies at fixed stellar mass. 
We show that the density-morphology-relation, i.e. the correlation between the 
quiescent fraction of satellites and the environmental density, starts to build 
up at $z \sim 2$ and is independent of the star formation properties of the
central galaxies. We conclude that environmental quenching is more important for satellite
galaxies than for centrals.
Our simulations also indicate that conformity is already in place at
$z=2$, where formation redshift and current star formation rate (SFR) of central and 
satellite galaxies already clearly correlate. Centrals with 
low SFRs tend to have formed earlier (at fixed stellar mass) while
centrals with high SFR formed later, with a typical formation redshift well
in agreement with current observations. 
However, we confirm the recent observations that the apparent number of 
satellites of spheroidal galaxies is significantly larger than for disc 
galaxies. This difference completely originates from the inclusion of 
companion galaxies, i.e. galaxies that do not sit in the potential 
minimum of a massive dark matter halo. Thus, due to the density-morphological-relation 
the number of satellites is not a good tracer for the halo mass, 
unless samples are restricted to the central galaxies of dark matter haloes only.  

\end{abstract}

\begin{keywords}
galaxies: evolution -- galaxies: formation -- galaxies: haloes -- galaxies: stellar content -- dark matter -- method: numerical
\end{keywords}


\section{Introduction}\label{sec:intro}
Observationally, dark halo masses for galaxies are difficult to measure. 
Indirect measurement methods are needed to infer the halo mass from observable
quantities, for example from hot X-ray halo gas \citep[e.g.][]{Kravtsov14,Gonzalez13},
weak galaxy-galaxy lensing \citep[e.g.][]{Mandelbaum06,Leauthaud12},
satellite velocities \citep[e.g.][]{Conroy07} or the number of
satellites \citep[e.g.][]{Wang12}, although discrepancies exist
between the results of the different methods \citep{Dutton10}.

Several observational studies find that the number of satellite
galaxies depends on the morphology of the host galaxy, namely that
early-type galaxies (ETGs) have more satellite galaxies than late-type
galaxies (LTGs) of the same stellar mass
\citep[e.g.][]{Wang12,Nierenberg12,Ruiz15}, leading to the conclusion
that ETGs live in more massive dark matter haloes than LTGs.
Furthermore, \citet{Lopez12} find that the ETGs have three times more
minor mergers than LTGs, with a basically constant rate since $z=1$,
which supports the idea that ETGs have more satellites than LTGs.
However, \citet{Liu11} do not find a different satellite fraction for
red and blue Milky-Way mass galaxies from SDSS.

The number of satellite galaxies is also correlated with the stellar
mass of the central galaxy \citep[e.g.][]{Bundy09,Nierenberg12,Wang12,Nierenberg16}, in that more
massive galaxies have more satellites. This implies that the dark
halo mass and the stellar mass of a host galaxy are also correlated,
which has been shown by observations \citep[e.g.][]{Mandelbaum06,Conroy07,More11,Gonzalez13,Kravtsov14,Velander14,Mandelbaum16,Wang16},
with indications for different relations for the different
morphological types \citep[e.g.][]{Mandelbaum06,More11,Velander14,Mandelbaum16}. 
The former is also supported by theoretical work using
semi-analytic modelling \citep{Guo13,Mitchell15}, abundance matching
\citep{Moster13,Behroozi13}), and simulations \citep[e.g.][]{Tonnesen15}. From their simulations, \citet{Tonnesen15} also
find that the stellar-to-halo-mass ratio is higher in over-dense
environments than in under-dense regions. However, observations do
not find a dependence of either the stellar-to-halo-mass relation
\citep{vanUitert16} nor the average halo mass of central galaxies
\citet{Brouwer16} on the environment.

For the environment, \citet{Dressler80} already found a relation
between galaxy number density and morphological type. He showed that
the ETGs are the dominant morphological type in galaxy cluster
environments, while the field environment is dominated by LTGs. Thus,
there must be a mechanism depending on the environment that leads to a
change in morphology. Over the years there have been several attempts
at explaining this phenomenon: In dense environments, the likelihood
for galaxy-galaxy merger is enhanced, especially in the group
environment \citep{Mamon92}, and it is well known that (dry) merger
events can lead to morphological transformations
\citep[e.g.][among other]{White78,Hernquist89,Springel00}.
However, in galaxy cluster environments the relative velocities of
galaxies are so large that the likelihood for a merger between two
arbitrary galaxies is decreasing again for all galaxies but the
central galaxy. While the merger likelihood in the densest
environments decreases again, there are other effects inside these
clusters that could lead to morphological transformations, for example
ram-pressure stripping \citep{Gunn72,Abadi99} and strangulation
\citep{Larson80}, by removing the gas from the galaxies and preventing
them from accreting new gas. These processes are generally summed up
as environmental quenching. While there is increasing observational
evidence for ongoing stripping processes in cluster environments
\citep{Abramson11,Vollmer12,Boissier12}, there is also evidence
that these mechanisms are only dominant for low mass galaxies and at
low redshifts
\citep{Peng10,Darvish16,Huertas16,Hirschmann16,Erfanianfar16}. At
higher redshifts ($z\geq 1$) and for more massive galaxies, they find
that internal quenching mechanisms like mass quenching or feedback
quenching from AGN and stellar winds seem to play a more important
role in quenching the star formation inside galaxies and thus
transform them into quiescent galaxies in addition to the known
channels of morphological transformations through merger events.

In this work we investigate the origin of the observed differences in
the satellite abundance around LTGs and ETGs of a given mass and study
the connection of this difference with the environment, using galaxies
from the Magneticum Pathfinder Simulations (Dolag et al., in
prep.,\citet{Hirschmann14a,Teklu15}). In Sec. \ref{sec:sim} we briefly
introduce the simulation and our method of selecting and classifying
the simulated galaxies. We demonstrate in Sec. \ref{sec:SMHM} that we
find a relation between the stellar and dark halo mass similar to the
observed one, and discuss possible sampling biases, while the origin
of the satellite abundance signal is investigated in \ref{sec:result}.
We investigate several properties connected to the SFRs of centrals
and satellites in Sec. \ref{sec:SFRs}, and explain the connection to
the morphology density relation.  Finally, we discuss and conclude our
results in Sec. \ref{sec:conclusion}.


\section{The Magneticum Simulation}\label{sec:sim}

The galaxies for this study were selected from the cosmological
hydrodynamical simulation set Magneticum Pathfinder\footnote{www.magneticum.org} (Dolag et al., in prep.).
The simulations were performed with an extended version of the
N-body/SPH code GADGET-3 which is an updated version of GADGET-2
\citep{Springel01a,Springel05}. They include various updates in the
formulation of SPH regarding the treatment of the viscosity and
the used kernels (see \citealp{Dolag05,Donnert2013,Beck15}). Our
simulations also include a wide range of complex baryonic physics such
as gas cooling and star formation \citep{SH03}, black hole seeding,
evolution and AGN feedback \citep{S05a,Fabjan10,Hirschmann14a} as well
as stellar evolution and metal enrichment \citep{Torna04,Torna07},
allowing each gas particle to form up to four star particles.  It also
follows the thermal conduction at 1/20th of the classical spitzer
value \citep{Spitzer62}, following \citet{Dolag04,Arth14}. More details can
be found in \citet{Teklu15}.

The initial conditions are based on a standard $\Lambda$CDM cosmology
with parameters according to the seven-year results of the Wilkinson
Microwave Anisotropy Probe (WMAP7) \citep{Komatsu11}. The Hubble
parameter is $h=0.704$ and the density parameters for matter, dark
energy and baryons are $\Omega_{M}=0.272$, $\Omega_{\Lambda}=0.728$
and $\Omega_{b}=0.0451$, respectively. We use a normalization of the
fluctuation amplitude at 8 Mpc of $\sigma_{8}=0.809$ and also include
the effects of Baryonic Acoustic Oscillations.

The Magneticum Pathfinder Simulations have already been successfully
used in a wide range of numerical studies, showing good agreement with
observational findings for the pressure profiles of the intra cluster
medium \citep{Planck2013pp,SPT2014pp}, the predicted Sunyaev Zeldovich
signal \citep{Dolag16}, the properties of the formed AGN population
\citep{Hirschmann14a,Steinborn15,Steinborn16}, the dynamical
properties of massive spheroidal galaxies \citep{Remus13,Remus16} and
the angular momentum properties of galaxies \citep{Teklu15}.

In this work we mainly use the medium-sized cosmological box {\it Box4} with a
volume of (48 Mpc/h)$^{3}$ at the {\it uhr} resolution level,
which initially contains a total of $2\cdot576^{3}$ particles (dark
matter and gas) with masses of $m_\mathrm{DM} = 3.6\cdot10^{7}
M_{\odot}/h$ and $m_\mathrm{gas} = 7.3\cdot10^{6} M_{\odot}/h$, having
a gravitational softening length of 1.4 kpc/h for dark matter and gas
particles and 0.7 kpc/h for star particles. Additionally,
we use the new cosmological box {\it Box3} with a larger volume of 
(128 Mpc/h)$^{3}$ at the same resolution level, evolved with a slightly
updated black hole treatment, as described by \citet{Steinborn15},
reaching a redshift of $z=2$.

For the identification of sub-halos we use a version of SUBFIND
\citep{Springel01b}, which is adapted to treat the baryonic component
\citep{Dolag09}. Halos are identified by SUBFIND \citep{Springel01b}
based on a standard Friends-of-Friends \citep{Davis85}. To 
calculate the virial radius ($R_\mathrm{vir}$) of halos, the
density contrast based on the top-hat model by \citet{Eke96} is used.
For comparison with some observations, we also use an over-density 
with respect to 200 times the critical density for calculating 
global properties when needed.


\subsection{Classification of Simulated Galaxies}\label{sec:sample}

From our dataset we extract all galaxies with stellar masses higher
than $10^{10} M_{\odot}$ from the simulation, regardless if they are the
central galaxy or a satellite of a larger halo. This limit was chosen to ensure a
sufficient resolution of the stellar content. This leads to a
total number of 2112 galaxies at $z=0$, see Table \ref{tab:no_gal}. 
We calculate the stellar specific
angular momentum $j$ for all galaxies, including all stars within a radius of
$5\cdot R_{1/2}$. We choose $R_{1/2}$ to be the stellar half-mass-radius
of all stars bound in the according subhaloe.

For a relatively simple but still efficient classification of the
galaxies we utilize the stellar mass vs. stellar angular momentum plane 
of galaxies. The position of galaxies on this plane was shown from both
observation \citep[e.g.][]{Fall83,Romanowsky12} and simulations \citep{Teklu15} 
to be a good indicator for the galaxy type. 

Following \citet{Teklu15}, for each galaxy we calculate the $b$-value 
\begin{equation}
  b =
  \mathrm{log}_{10}\left(\frac{j}{\mathrm{kpc}~\mathrm{km/s}}\right) -
  \frac{2}{3}\mathrm{log}_{10}\left(\frac{M}{M_{\odot}}\right),
\end{equation}
which is the $y$-intercept of the linear relation $f(x)=ax+b$ in the
log-log of the stellar mass vs. stellar angular momentum plane.
At $z=0$, galaxies with $b > -4.35$ are classified as disc galaxies,
while galaxies with $b<-4.73$ are classified as ellipticals. Everything
in between are labeled as intermediates. For higher redshifts, 
we adopt the theoretically expected scaling derived by \citet{Obreschkow15}, 
which was shown to be an excellent match for both disc and elliptical galaxies 
in the simulations at $z=2$ by \citet{Teklu16}.

In Table \ref{tab:no_gal} we list the resulting number of galaxies classified 
into spheroids, intermediates, and discs at four different redshifts ($z=0$, 
$z=0.5$, $z=1$ and $z=2$). While clear spheroids are getting less
frequent at higher redshift, the fractions of disc and intermediate
galaxies increase with higher redshift.

\begin{table}
\caption{The number of all host galaxies with stellar masses higher than $10^{10} M_{\odot}$ at four different redshifts}
\label{tab:no_gal}
\centering
 \begin{tabular}{l || l l l | l l }
 \hline \hline
  redshift & $N_\mathrm{spheroid}$ & $N_\mathrm{interm.}$ & $N_\mathrm{disc}$ & $N_\mathrm{quiescent}$ & $N_\mathrm{SF}$ \\ \hline
  0 & 656 & 760 & 696 & 1763 & 349\\
  0.5 & 549 & 861 & 815 & 1697 & 528\\
  1 & 419 & 934 & 857 & 1374 & 836\\
  2 & 246 & 869 & 605 & 180 & 1440\\
 \hline  
 \end{tabular}
\end{table}

As observations often utilize the specific star formation rate (sSFR) as an indicator for 
the morphology of galaxies, we also split our sample into quiescent and star
forming galaxies. Following the classification by \citet{Franx08}, 
we define quiescent galaxies to have a sSFR smaller than $0.3/t_\mathrm{Hubble}$, 
where $t_\mathrm{Hubble}=1/H(z)$ is the Hubble time, while the rest is classified
as star-forming. The resulting number of quiescent and star-forming galaxies 
for different redshifts is listed in Table \ref{tab:no_gal}.


\section{The Relation between the Stellar and Halo Mass of Central Galaxies}\label{sec:SMHM}

\begin{figure*}
    \begin{centering}
    \includegraphics[width=0.7\textwidth,clip=true]{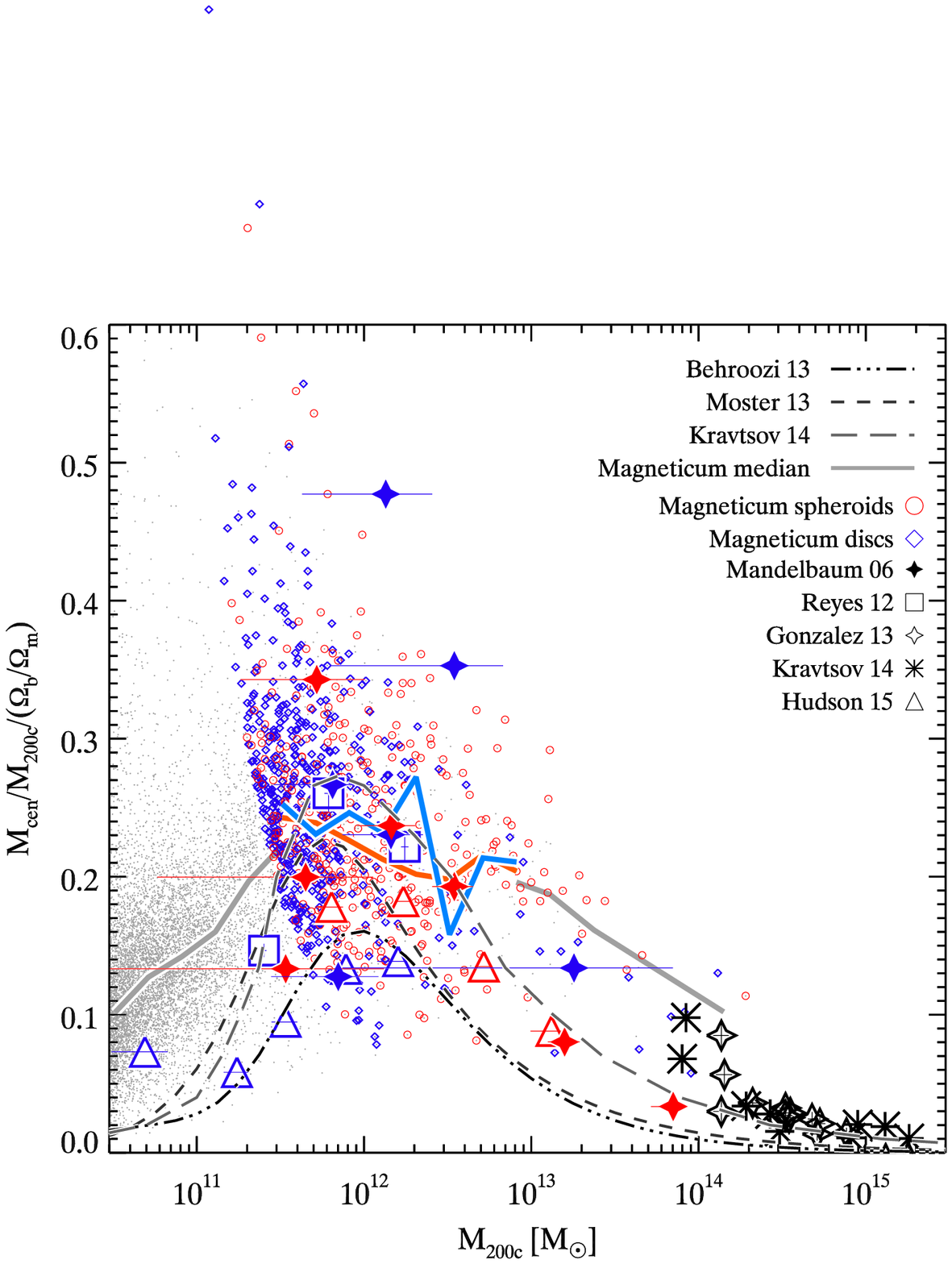}
    \caption{The stellar mass fraction of central galaxies with respect
to critical halo mass $M_{200\mathrm{c}}$ normalized by the global, cosmological
baryon fraction $\Omega_{b}/\Omega_{m}$, as a function of the halo mass 
$M_{200\mathrm{c}}$. There is no difference between spheroids (red circles) and discs (blue diamonds). }
    \label{fig:m200c_mcd}
  \end{centering}
\end{figure*}

Over the past years several studies investigated the stellar-to-halo-mass 
relation \citep[e.g.][]{Mandelbaum06,Behroozi10,Dutton10,Moster10,Wake11,Munshi13,Kravtsov14,Brook14,Shankar14,Rodriguez15,vanUitert16,Tinker16}. This relation, applicable only for galaxies 
at the center of the underlying dark matter halo, can give important insight 
into the galaxy formation process. One interesting question is, for example,
if galaxies of identical stellar mass but different morphological type are
hosted by dark matter halos of different masses. Another interesting aspect is,
that the efficiency with which galaxies are converting their cosmological baryon
reservoir into stars, turned out to depend on halo mass, where galaxies residing
in dark matter halos of approximately $10^{12}M_\odot$ appear to be most efficient. 

To investigate this aspect, we show the baryon conversion efficiency as function of halo mass
for our central galaxies in Fig. \ref{fig:m200c_mcd}, where we show the elliptical galaxies
as red circles and the disc galaxies as blue diamonds. Intermediate galaxies and
galaxies below our mass cut are shown as gray dots. The gray solid lines mark the 
median of all galaxies, while the red and blue solid lines show the median of the
ellipticals and discs, respectively. For comparison, we over-plot data
points from observations by \citet{Mandelbaum06} (filled stars),
\citet{Reyes12} (squares), \citet{Gonzalez13} (open stars),
\citet{Kravtsov14} (asterisks) and \citet{Hudson15} (triangles), where
the colours red and blue represent samples of elliptical/early-type
galaxies and disc/late-type galaxies, respectively.
Our simulated galaxies agree qualitatively with the different observations, 
which show a wide spread. In both simulations and observations the 
differences for different galaxy types are only marginal. The differences between
our simulations and the observations are most prominent at the low and very high-mass end. 
They are driven by too inefficient stellar feedback for the low-mass haloes
and too inefficient black hole feedback at the high-mass end.

Only at the very high-mass end, X-ray observations allow to infer unambiguously
the underlying dark matter potential of individual galaxies \citep{Kravtsov14,Gonzalez13}. 
In addition, the outer stellar components of these massive galaxies can be measured and 
their total stellar mass can be inferred, pushing the observations closer to our 
simulation result. We also plot the curves obtained with abundance matching by 
\citet{Moster13} (dashed line),  \citet{Behroozi13} (dash-dotted line), and \citet{Kravtsov14}
(solid line). Especially, the results obtained by \citet{Kravtsov14} resemble very
closely the simulation result.

\begin{figure}
    \begin{centering}
    \includegraphics[width=0.45\textwidth,clip=true]{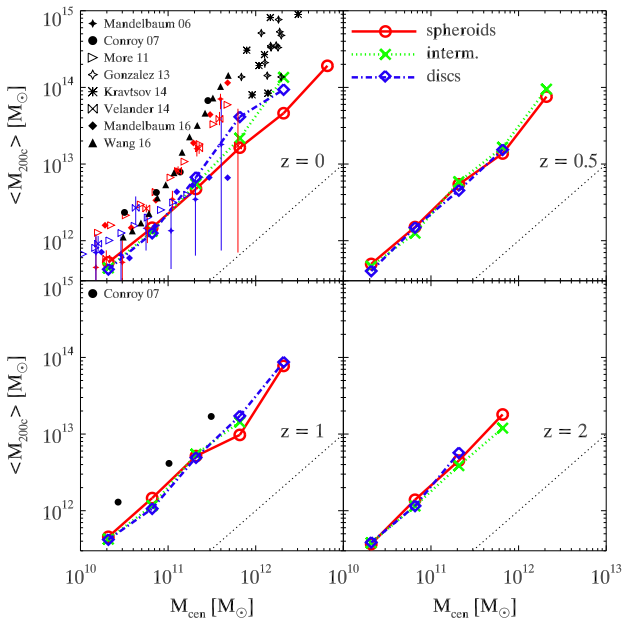}
    \caption{The average critical halo mass $M_\mathrm{200c}$ of a
      halo as a function of the stellar mass of the central galaxy for
      different redshifts. }
    \label{fig:mvir_ms_z}
  \end{centering}
\end{figure}

A direct comparison between the total mass of the halo and the stellar mass 
of the central galaxy and its evolution with time is shown in Fig. \ref{fig:mvir_ms_z}.
Red, blue and green lines represent the relation for the ellipticals, discs, and intermediates
in the simulation, respectively. At redshift $z=0$ (upper left panel) we add data points 
from \citet{Mandelbaum06},
\citet{Conroy07}, \citet{More11}, \citet{Gonzalez13},
\citet{Kravtsov14}, \citet{Velander14}, \citet{Mandelbaum16} and
\citet{Wang16}. Black symbols represent those studies that do not 
distinguish between galaxy types, while red symbols represent
elliptical/red/early-type galaxy samples and blue symbols represent
spiral/blue/late-type galaxy samples.

At masses below $2\times10^{11}M_\odot$ the simulations overall lie within the large spread
covered by the observations. Although the simulations best resemble the results by 
\citet{Mandelbaum06,Mandelbaum16}, they generally predict slightly higher stellar mass at given 
halo masses than most observations. This difference increases with stellar mass. However,
at the very high-mass end, where observations can properly include the outermost parts of the galaxies
in form of the ICL component \citep{Kravtsov14,Gonzalez13}, the observations lie again closer to our 
results. Therefore, this systematic upturn seen in the observations 
could be related to the effect of the contribution of the outer halo to the total stellar mass.

Using observational data from the SDSS \citet{Mandelbaum06} and \citet{More11}
concluded that the halo mass is independent of the morphology at a fixed
stellar mass below $10^{11}$ and $2 \cdot 10^{10.5} M_{\odot}$, respectively, while both found that red/elliptical galaxies 
reside in more massive haloes at higher stellar masses. 
They argue that this halo mass is likely to reflect the mass of the cluster/group in
which more massive ETGs live in. 
\citet{Mandelbaum16} find that passive galaxies with stellar masses between $10^{10.3}$ and $10^{11.6} M_\odot$ reside in more massive halos than their star-forming counterparts.
While our simulations also do not show
any differences for the host halo masses at low stellar masses for ellipticals 
and discs, we cannot support this conclusion that dark matter haloes of spheroids 
are more massive than those of disc galaxies with comparable stellar masses. 
In contrast, we even find that disc galaxies with stellar masses above $2\times10^{11}M_\odot$ 
live in more massive haloes than spheroids of the same stellar mass, however, this
could be due to low number statistics, since at high stellar masses there are only 
few discs in our simulation. 
 
This discrepancy vanishes at higher redshift, where we do not find
any differences between the ellipticals and discs, even at large
stellar masses. Furthermore, we do not find any evolution of the
stellar-to-halo-mass relation with redshift at all, in agreement
with observations
using satellite kinematics by \citet{Conroy07}, who find that the the
stellar to halo mass ratio does not evolve between $z \sim 0$ and $z
\sim 1$ for their sample of observed host galaxies below
$M_\mathrm{star} \approx 1.5\times10^{11} M_{\odot}$. This
is partly in agreement with \citet{Hudson15} who find no significant redshift
evolution for their sample of blue galaxies. However, they
observe a time evolution of the relation for their
red galaxies.   


\begin{figure*}
  \begin{centering}
    \includegraphics[width=0.9\textwidth,clip=true]{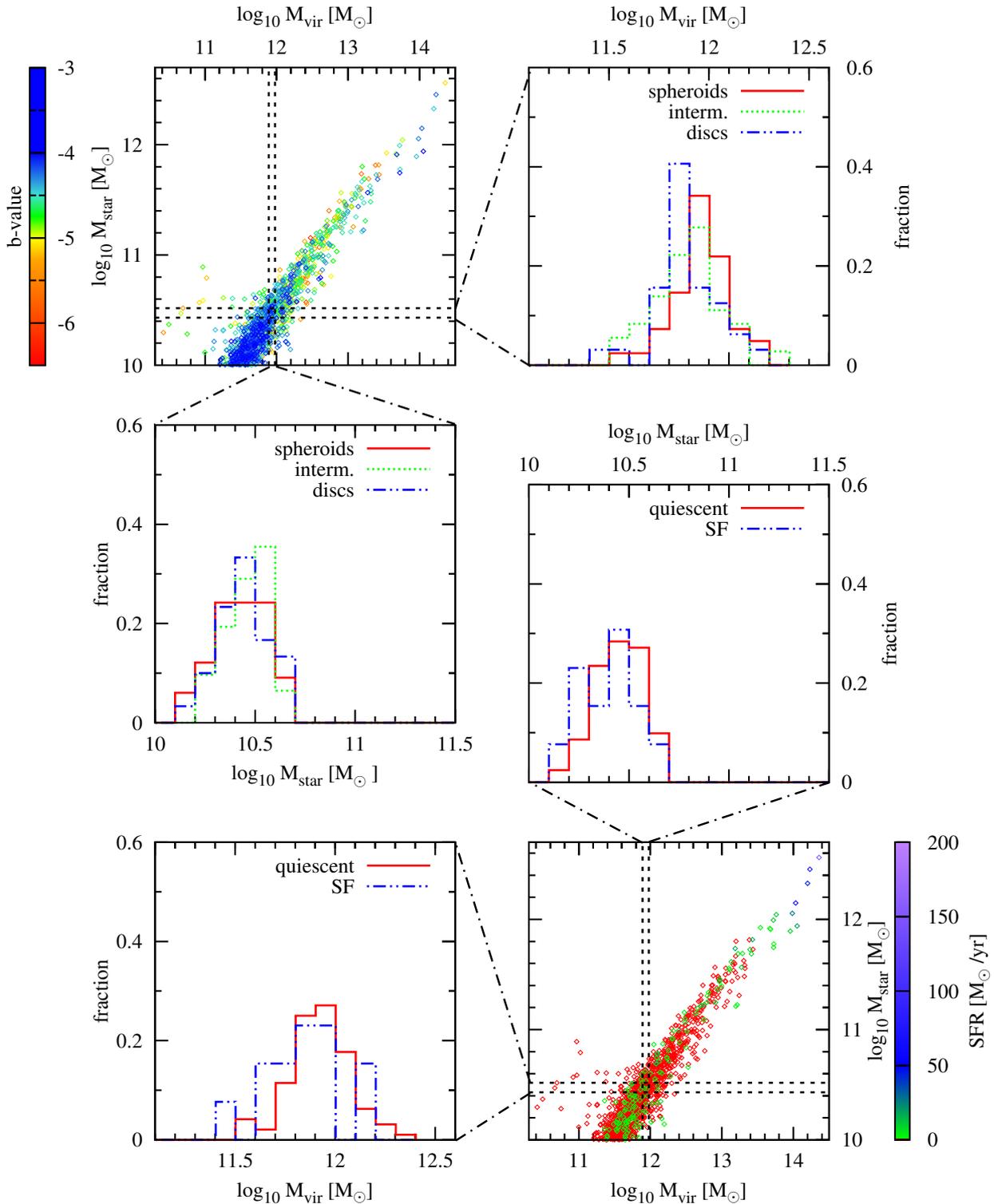}
    \caption{The stellar to halo mass relation for the centrals in Box4 at redshift $z=0$. \textit{Top left:} the virial mass against the stellar mass of individual centrals, colour-coded by the $b$-value; \textit{top right:} the fraction of centrals with $M_\mathrm{star} \in [2.7 \cdot 10^{10},3.3 \cdot 10^{10}]$ in bins of $M_\mathrm{vir}$, divided into spheroids, intermediates and discs; \textit{middle left:} the fraction of centrals with $M_\mathrm{vir} \in [7.8\cdot 10^{11},9.5 \cdot 10^{11}]$ in bins of $M_\mathrm{star}$. 
    \textit{Bottom right:} the virial mass against the stellar mass of individual centrals, colour-coded by the SFR; \textit{bottom right:} the fraction of centrals with $M_\mathrm{star} \in [2.7 \cdot 10^{10},3.3 \cdot 10^{10}]$ in bins of $M_\mathrm{vir}$ divided into quiescent and star-forming; \textit{middle right:} the fraction of centrals with $M_\mathrm{vir} \in [7.8\cdot 10^{11},9.5 \cdot 10^{11}]$ in bins of $M_\mathrm{star}$. }
    \label{fig:cen_ms_mv_b_qu}
  \end{centering}
\end{figure*}  


In Fig. \ref{fig:cen_ms_mv_b_qu} we take a closer
look at the stellar-to-halo-mass relation at a fixed stellar and
virial mass regarding the different classifications, i.e. the classification according to the
kinematical $b$-value and that according to the sSFR, of the
central galaxies at $z=0$. The upper left panel shows the relation
for individual central galaxies colour-coded by the dynamical
classification parameter $b$-value. There is no observable trend 
with the $b$-value. The upper right panel shows the
distribution of the virial masses of the centrals with stellar masses
in the small range between $2.7 \cdot 10^{10} M_\odot$ and $3.3 \cdot
10^{10} M_\odot$.  In agreement with Fig. \ref{fig:mvir_ms_z}, we find
that the spheroidal galaxies peak at higher virial masses than the
discs. When looking at the distributions of centrals of virial masses
in the range between $7.8 \cdot 10^{11} M_\odot$ and $9.5 \cdot
10^{11} M_\odot$ (middle left panel) we do not find a difference in
the distribution.

The right bottom panel shows the relation for the
galaxies colour-coded by the SFR, where centrals with
SFRs $<1 M_{\odot}/yr$ are red. As on the upper left panel there
is no visible trend between the two quantities and the SFR. 
In contrast to the result where we have taken the $b$-value (upper right),
the same galaxy population in the same small stellar mass
bin shows no difference in the distribution in the virial masses of
quiescent and star-forming centrals (bottom left).
There is also no difference in the distribution of the
stellar masses at fixed virial mass (middle right panel),
which is in agreement with the results from the $b$-value classification. 

This result demonstrates that different classifications can
lead to different conclusions regarding the relation between the
virial mass and the stellar mass and the morphology. Furthermore,
we find a dependence on the mass range: As shown in 
Fig. \ref{fig:mvir_ms_z}, at higher mass this trend even reverses.
Additionally, we clearly see that a different behaviour occurs if 
the distribution at fixed stellar mass (where in this case we see a
different behaviour) is considered, or the distribution at fixed
virial mass (where in this case we do not see a different behaviour) 
is taken.  

 
\begin{figure}
  \begin{centering}
    \includegraphics[width=0.47\textwidth,clip=true]{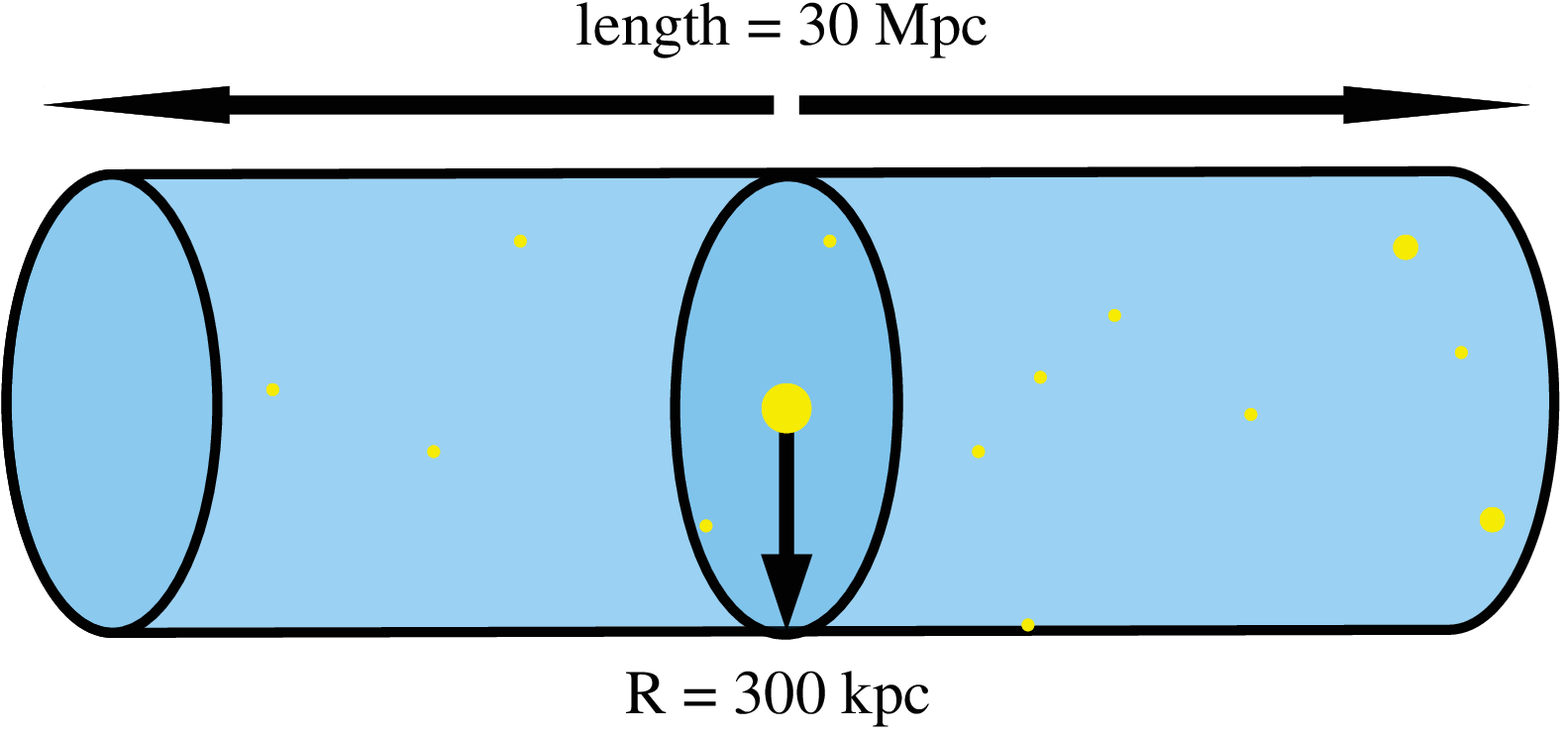}
    \includegraphics[width=0.47\textwidth,clip=true]{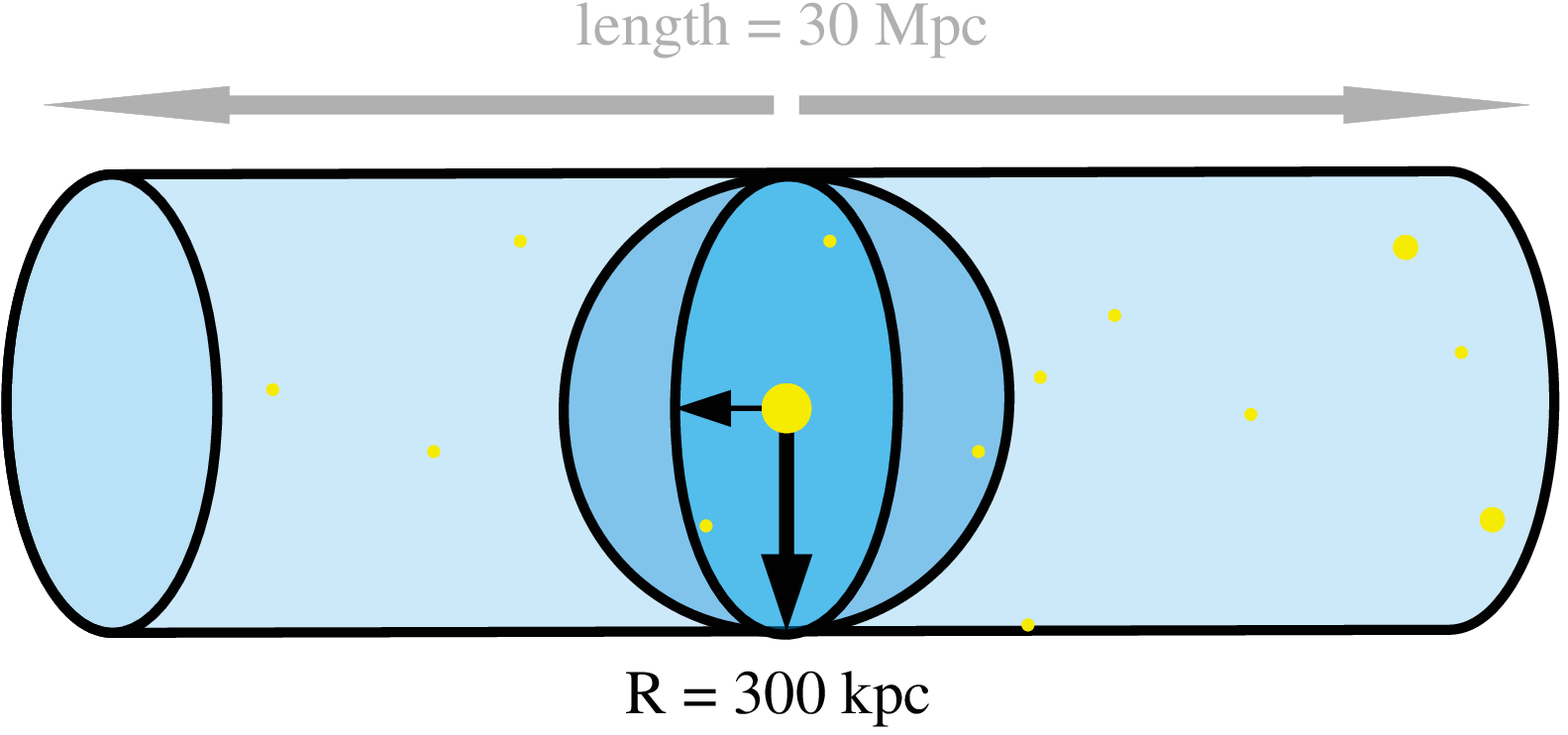}
    \caption{Selecting galaxies in different ways: The upper figure shows illustrates Method 1, where we select satellites within a cylinder of radius $r=300$kpc and a length of 30 Mpc. 
    The lower figure illustrates the second method, where we select all satellites within a sphere of 300kpc.}
    \label{fig:count_schema_geometry}
  \end{centering}
\end{figure}  


\section{The Abundance of Satellites}\label{sec:result}

In a recent observational study, \citet{Ruiz15} 
tried to infer the dark matter halo masses of galaxies based on the abundance
of their satellites. They found a systematically decreasing number of
satellites around galaxies along the Hubble sequence. 
As in the observations, we target all galaxies above a given mass threshold
in our simulations, regardless of them being centrals or not, and assign all
surrounding galaxies within a mass ratio of $M_\mathrm{sat}/M_\mathrm{host} \in [0.01,1]$
to be their apparent satellite galaxies. 

By different ways of counting, as shown in Fig. \ref{fig:count_schema_geometry}, 
we can estimate the contribution of projection effects
by performing our selection in two different ways: 
First, we select apparent satellite galaxies within a cylinder of radius $r$ and with a length $l$, as usually done in observations; 
Second, we select apparent satellite galaxies within a sphere of radius $r$ around a target galaxy.

To obtain a meaningfully statistical number of galaxies we had to choose a mass threshold 
of $M_\mathrm{star} \ge 10^{10}M_\odot$, which is lower than the one used by \citet{Ruiz15}. However, 
in respect to $r$ and $l$ we follow the observations by \citet{Wang12} and \citet{Ruiz15},
namely by choosing a radius $r = 300 \mathrm{kpc}$ and a length $l = +/- 15 \mathrm{Mpc}$ 
along the line of sight, which correspond to the redshift interval of $|\Delta z| < 1000 \mathrm{km s^{-1}}$. 


\subsection{Comparison of the Satellite Number with Recent Observations}\label{sec:obs}

In the following section we compare the abundance of inferred satellites around
the selected sample of galaxies in our simulation with the results from
\citet{Ruiz15}. Therefore, we use the first selection method,
where the apparent satellites are counted within a cylinder to calculate 
the number of inferred satellites around massive galaxies.


\begin{figure}
  \begin{centering}
    \includegraphics[width=0.45\textwidth,clip=true]{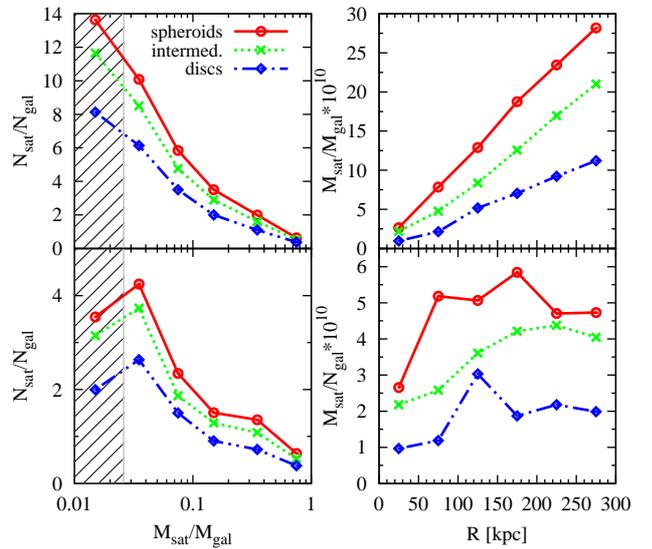}
    \caption{\textit{Upper Left:} The mass fraction of satellites with respect to the stellar mass of the host galaxies against the cumulative number of satellites per host galaxy in a cylinder of radius $r = 300 \mathrm{kpc}$ and a length of $30 \mathrm{Mpc}$.  
    \textit{Lower left:} The differential number of satellites per host galaxy. 
    \textit{Upper right:} The cumulative radial distribution of the satellite mass around host galaxies. 
    \textit{Lower right:} The differential distribution of the satellite mass along the radius.}
    \label{fig:m_n_r_m}
  \end{centering}
\end{figure}

In the left panels of Fig. \ref{fig:m_n_r_m} we show the mean number of inferred
satellites as function of their stellar mass ratio to the host galaxy at $z=0$.  
The upper panel shows the (anti-)cumulative fraction, while on the lower 
panel the differential fraction is shown. 
The shaded area indicates the completeness of our sample, assuming that 
galaxies with more than 100 star particles are always numerically resolved. 
The number of satellites for spheroidal host galaxies is shown in red, while 
the according numbers for disc host galaxies are shown in blue.
In agreement with observations by \citet{Ruiz15} (see also e.g., \citet{Wang14}), 
we find that spheroidal host galaxies have more inferred satellites than disc galaxies.\footnote{The differences 
in the absolute number of apparent satellites between the simulations and the observations is driven by the different 
masses in the two samples.} 


The radial distribution of the apparent satellite mass around the host at redshift $z=0$
is shown in the right panels of Fig. \ref{fig:m_n_r_m}, where
the upper one shows the cumulative number, while the
lower one shows the differential mass distribution.
For all galaxy types the mass in inferred satellites rises towards the outer regions. 
We find that spheroids are surrounded by $\propto 3$ times the mass in inferred satellites
compared to discs and $\propto 1.5$ times compared to intermediates.
This tendency agrees qualitatively well with the observations by \citet{Ruiz15}.

\begin{figure}
  \begin{centering}
    \includegraphics[width=0.48\textwidth,clip=true]{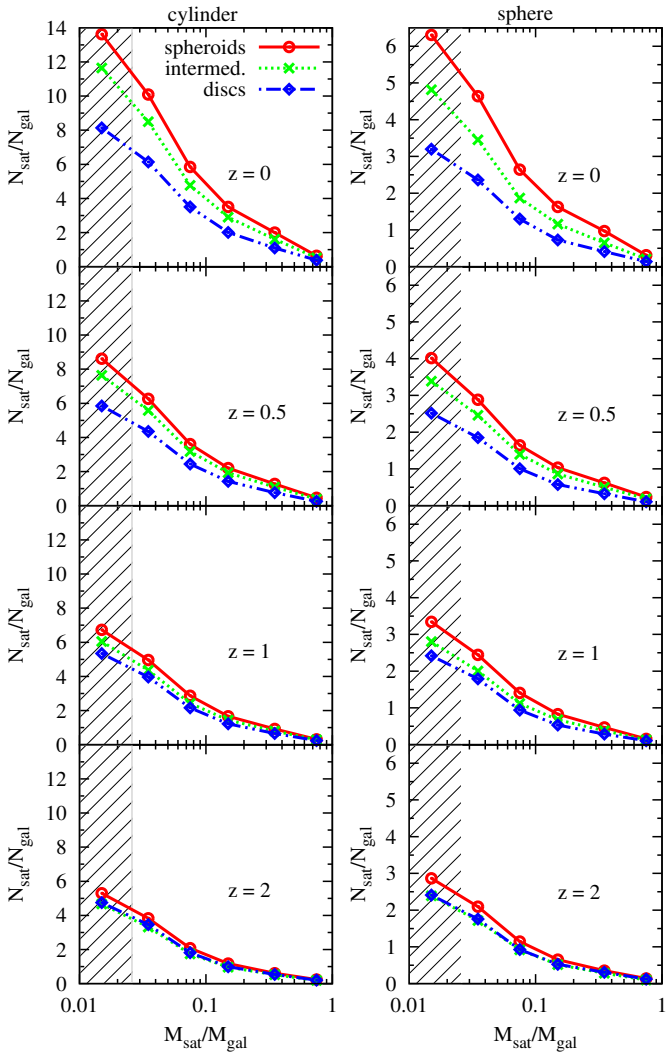}
    \caption{The cumulative number of satellites per host galaxy at different redshifts for all host galaxies. The \textit{left} column shows the results for a cylinder, while for the \textit{right} column we collect satellites within a sphere of $r=300 \mathrm{kpc}$ (co-moving).
    The split-up of the distribution is visible for all redshifts, but it becomes smaller with increasing redshift. Be aware: y-axes have different scales!}
    \label{fig:m_n_cyl_sph}
  \end{centering}
\end{figure}

\subsection{Abundance of Satellites for $z > 0$}\label{sec:high_z}

A more reasonable selection (which obviously is not possible in observations)
is the second selection method, where the apparent satellites are counted 
within a sphere around the host galaxy, which is more closely related to
the physical satellite population and in principal better allows to
study the evolution of the satellite population. 
As shown in Fig. \ref{fig:m_n_cyl_sph}, the different number of inferred 
satellite galaxies around spheroidal galaxies compared to disc galaxies
is already present at higher redshift, independent of the counting method.   
Interestingly, the relative strength of the signal is even larger when
counting apparent satellite galaxies within spheres compared to cylinders, indicating 
that the immediate environment contributes the most.

The difference of the number of inferred satellites between the spheroids and discs
becomes smaller with increasing redshift, due to the decreasing number
of inferred satellites around spheroids, while the number of inferred satellites stays 
constant for disc galaxies. Our results are in agreement
with studies by \citet{Marmol12}, \citet{Nierenberg12} and
\citet{Nierenberg16}, who find no
significant redshift evolution of the number of apparent satellites for 
the redshift range between $0.2 < z < 2$, $0.1 < z <
0.8$ and $0.1 < z < 1.5$, respectively. They also
mention that the difference in the abundance of inferred satellites between
spheroids and disc-like galaxies is more prominent at lower redshift;
This likely is an effect of clustering of spheroidal galaxies, which
is more important at lower redshifts \citep[see][]{Marmol12}.

In this context, we want to mention that it is important to keep in mind, 
in which units the distances are measured, i.e. in physical or co-moving,
which becomes more and more important with increasing redshift. 

Since it is common to use physical units in observations, we show the
result for redshift $z=2$ with physical units in Fig. \ref{fig:m_n_phys} 
in the appendix. However, throughout this study we use co-moving units
to separate between the expansion of the universe and the real growth 
of structures.


\subsection{Radial Distribution of Satellites}\label{sec:rad_sat}

\begin{figure}
    \begin{centering}
    \includegraphics[width=0.45\textwidth,clip=true]{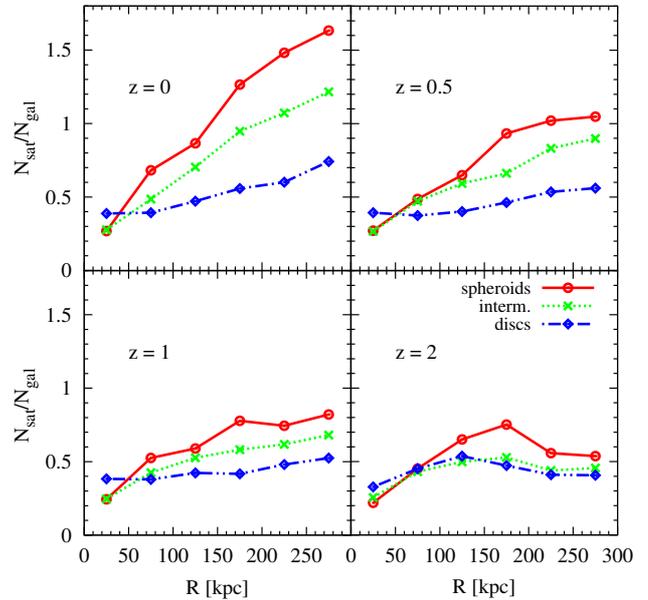}
    \caption{The differential number of satellites per host galaxy depending on the radius of the host galaxies for different redshifts in co-moving units. With increasing redshift the number of satellites decreases, especially for the spheroids. }
    \label{fig:n_r_z}
  \end{centering}
\end{figure}

In Fig. \ref{fig:n_r_z} we study how many inferred satellites within a sphere are located in
different bins of the radius for four different redshifts.
We clearly see that
spheroidal galaxies (red circles) have more apparent satellites at all radii
than disc galaxies (blue diamonds), except for the inner most region.
This is in agreement with studies by \citet{Wang14} who find similar
results for different mass ranges of isolated galaxies from the SDSS.
At low redshift (upper left) there is an almost linear distribution of
inferred satellites surrounding spheroids, increasing from inside out.  While
in the inner parts the number of inferred satellites is relatively constant
towards higher redshifts, in the outer region the number of inferred satellites
decreases with increasing redshift. For intermediates a similar but
weaker trend can be seen. In contrast to the spheroids and
intermediates, disc galaxies do not have a much ascending slope; the
apparent satellites are evenly distributed along the radius and there is no
evolution with redshift. 

\begin{table}
\caption{The number of central and companion galaxies with stellar masses higher than $10^{10} M_{\odot}$ at four different redshifts}
\label{tab:no_comp_cent}
\centering
 \begin{tabular}{l l l l}
 \hline \hline
 centrals & & & \\ 
 \hline \hline
  redshift & $N_\mathrm{spheroid}$ & $N_\mathrm{interm.}$ & $N_\mathrm{disc}$ \\ \hline
  0 & 396 & 442 & 481 \\
  0.5 & 344  & 539  & 561  \\
  1 & 285  & 630  & 629  \\
  2 & 179  & 663  & 463   \\
  \hline \hline
 companions & & & \\ 
 \hline \hline
  redshift & $N_\mathrm{spheroid}$ & $N_\mathrm{interm.}$ & $N_\mathrm{disc}$ \\ \hline
  0 & 260 & 318 & 215 \\
  0.5 & 205  & 322  & 254  \\
  1 & 134  & 304  & 228  \\
  2 & 67  & 206  & 142   \\
 \hline  
 \end{tabular}
\end{table}

 
\begin{figure}
  \begin{centering}
    \includegraphics[width=0.47\textwidth,clip=true]{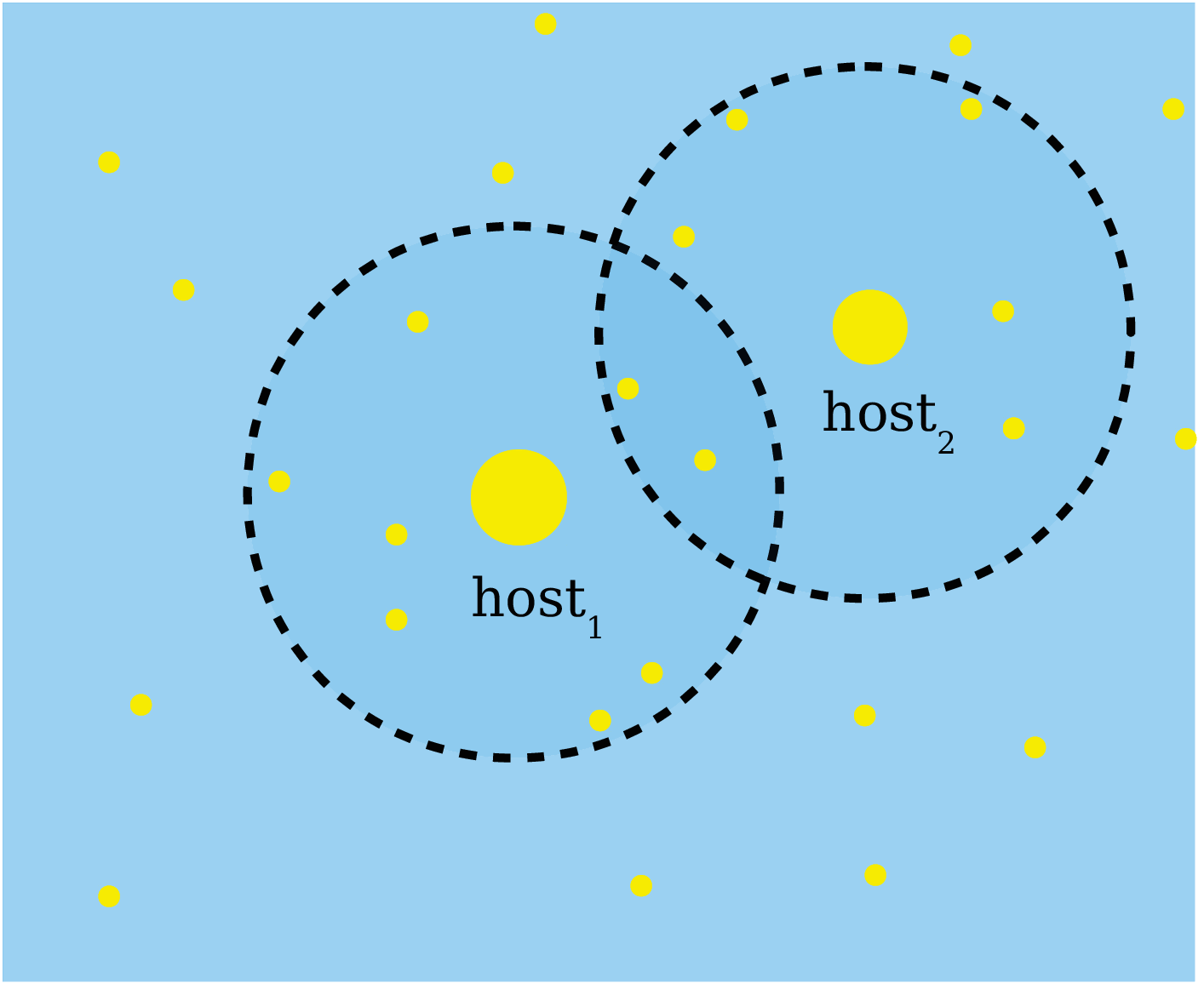}
    \includegraphics[width=0.47\textwidth,clip=true]{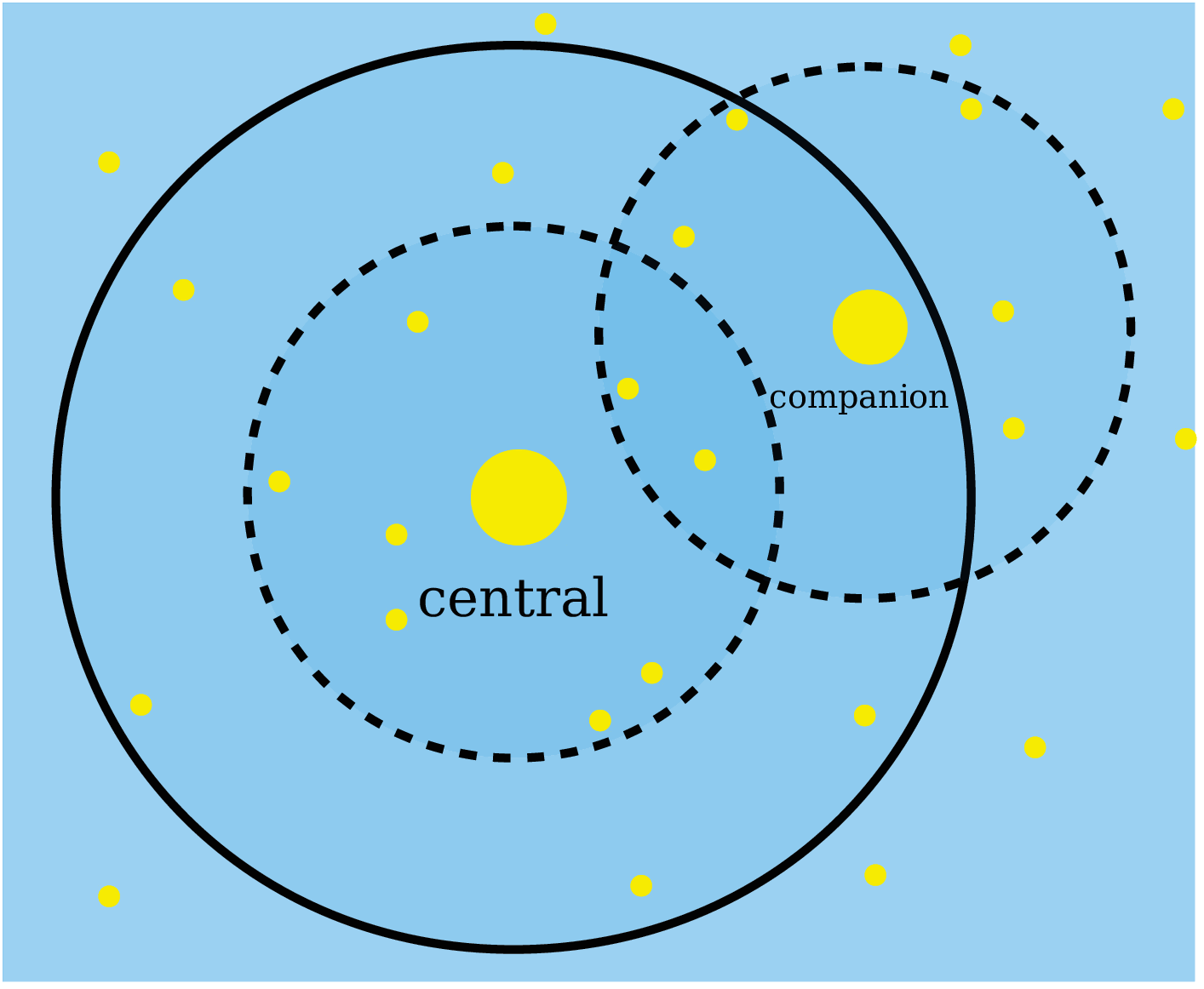}
    \caption{Selecting host galaxies in different ways. The upper figure illustrates the selection of satellites for ``all galaxies'': Every galaxy more massive than $10^{10} M_\mathrm{star}/M_{\odot}$ is counted as host galaxy. 
    The lower scheme shows the split-up into central and companion galaxies. While host$_1$ in the upper figure is a central, i.e. the galaxy in the potential minimum of the dark matter halo (black solid circle), host$_2$ from the upper figure is a companion, i.e. a large satellites within the same dark matter halo.}
    \label{fig:count_schema_type}
  \end{centering}
\end{figure}

\begin{figure*}
  \begin{centering}
    \includegraphics[width=0.9\textwidth,clip=true]{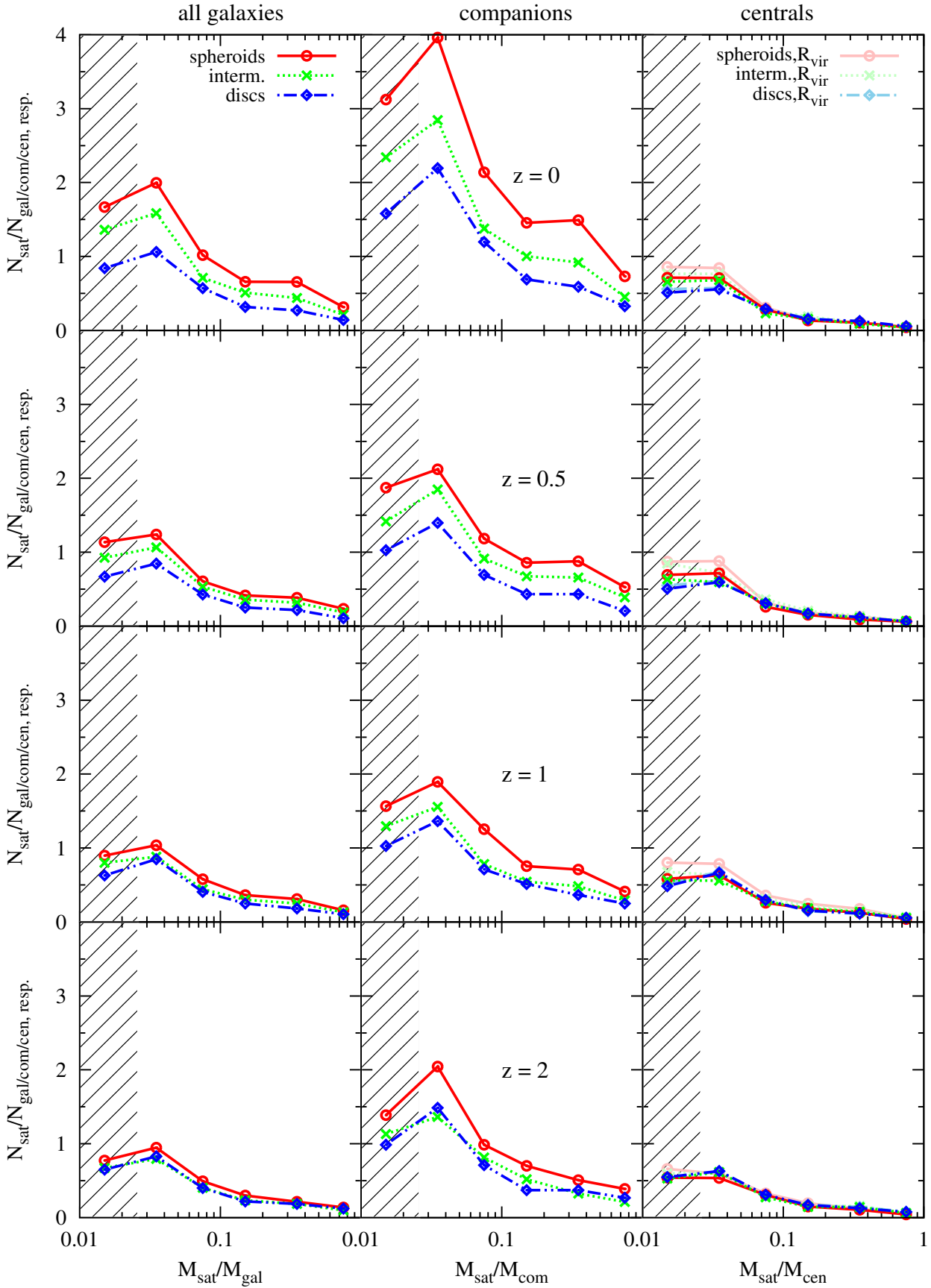}
    \caption{The differential number of satellites per host galaxy at different redshifts (rows) for all host galaxies (left panels), companions (middle panels) and central galaxies (right panels) in co-moving units. 
    The normal colours show satellites within $r=300$kpc, and the pale colours show satellites within $R_\mathrm{vir}$.
    The split-up of the distribution is mainly caused by the companions, since the centrals do not exhibit such a split-up. The split-up in the distribution for the companions is present at all redshifts.}
    \label{fig:m_n_com_cen}
  \end{centering}
\end{figure*}

\subsection{Comparison of Companions and Centrals}\label{sec:comp_cent}

To better understand the origin of the observed signal, we can now split our sample 
of target galaxies in {\it centrals} (which are, as discussed before, at the center of a 
dark matter halo) and {\it companions}, which are by themselves already parts of a 
larger dark matter halo. Therefore, we count every galaxy from the host galaxy sample as a
central galaxy if it resides in the potential minimum of a dark matter
halo. All remaining galaxies, which sit within the virial radius of a
central galaxy, are considered as companions. This different way of counting compared to
what was done before is illustrated in Fig. \ref{fig:count_schema_type}.
The resulting numbers of companion and central galaxies classified into spheroids,
intermediates and discs at four different redshifts are listed in
Table \ref{tab:no_comp_cent}. 

Fig. \ref{fig:m_n_com_cen} shows the number of satellite galaxies inferred for
all host galaxies (left column) compared to the number of satellite galaxies 
for the sample split into {\it centrals} (right panel) and {\it companions} 
(middle panel).
We see a clear difference in the number of apparent satellites: While there is a clear
split-up between spheroidal and disc galaxies for the host and companion samples,
there is no such difference visible for the centrals.
 
Thus, we clearly see that the companions within the sample are responsible 
for the split-up seen for the sample of all host galaxies.  
This can be explained by the fact that {\it companions} are by definition
in dense regions and due to the morphology-density-relation preferentially
are spheroidal galaxies. Therefore, we conclude that the observed difference
in the apparent number of inferred satellite galaxies around spheroidal and disc
galaxies is entirely driven by the morphology-density-relation and not by
different underlying halo masses.   

This also explains the weakening of the signal with increasing redshift,
as shown in the lower panels of Fig. \ref{fig:m_n_com_cen} and in Fig. \ref{fig:m_n_cyl_sph}. 
This is caused by the lack of massive companion
galaxies at higher redshifts as can clearly be seen from Table
\ref{tab:no_comp_cent} and the middle and right panels of
Fig. \ref{fig:m_n_com_cen}. Thus, the split-up of the distributions of
companions is visible at all redshifts but less pronounced at higher
redshift. 

This is also supported by the radial distribution of satellites as
previously shown in Fig. \ref{fig:n_r_z} (and in Fig. \ref{fig:r_n_cen_com}, separately for centrals and companions): The spheroidal host
galaxies have more satellites than the discs and the number of these
satellites increases towards larger radii, again pointing towards
the density-morphology-relation as origin of the split up.


\subsection{Average Number of Satellites in Dependence of the Virial Mass}\label{sec:rad_sat}

\begin{figure}
    \begin{centering}
    \includegraphics[width=0.45\textwidth,clip=true]{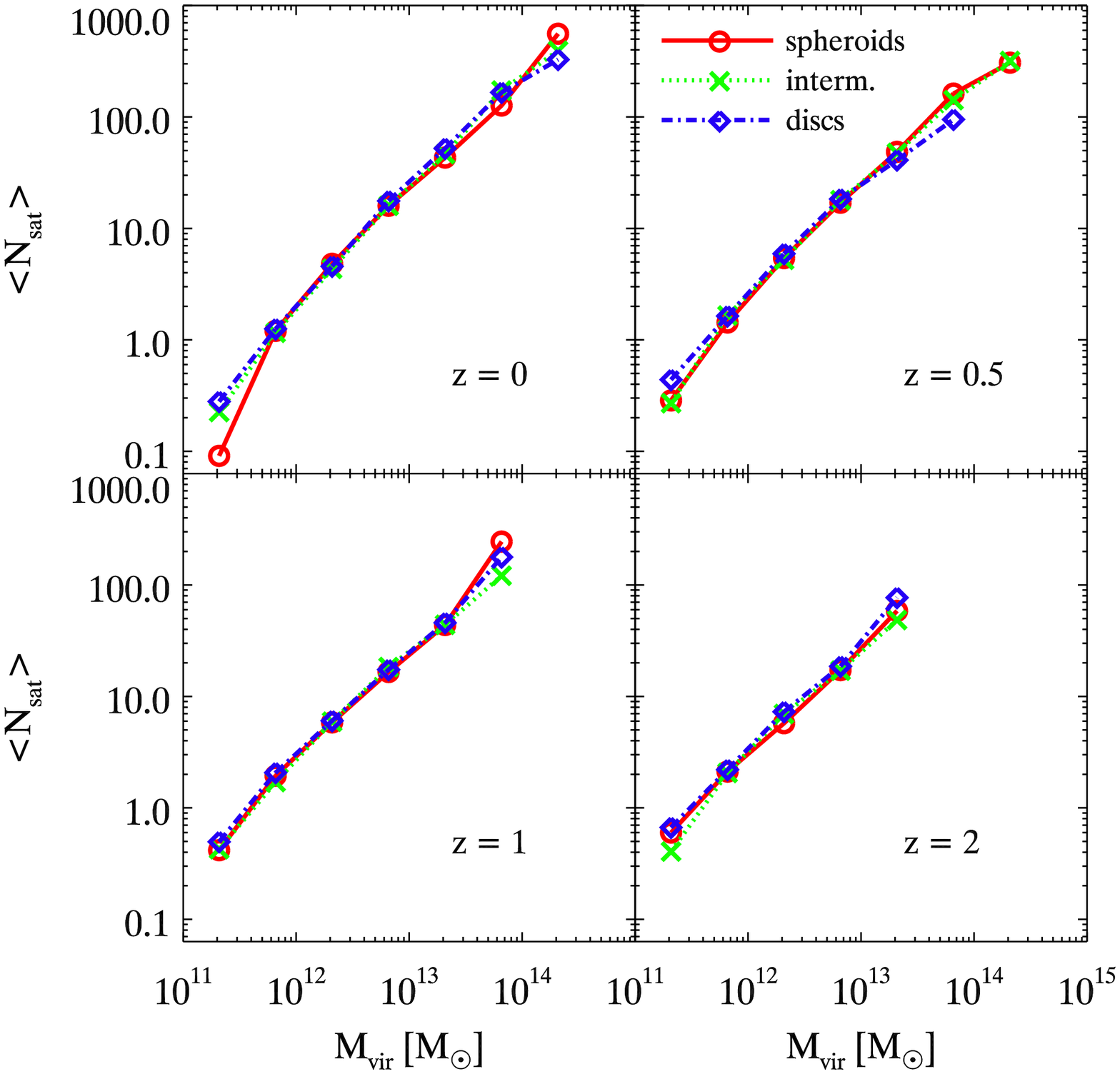}
    \caption{The average number of satellites with stellar masses larger than $10^{8} M_{\odot}$ around centrals, within the virial radius in bins of the virial mass for different redshifts. }
    \label{fig:mvir_nssat_z}
  \end{centering}
\end{figure}

In order to see if one can directly infer the halo mass of a central
galaxy from the number of the satellites, we show in
Fig. \ref{fig:mvir_nssat_z} the number of satellites averaged for the
central galaxies according to their virial mass. Here, we include all
satellites with stellar masses higher than $10^{8} M_{\odot}$ that
reside within the virial radius of their central galaxy. We find
that the number of satellites is directly proportional to the virial mass of the
central galaxy, where the central galaxies which reside in halos with
higher masses have more satellites than those residing in low-mass
halos. There is no difference for galaxies with different
morphological types. In addition, we do not find an evolution with
redshift. 

The independence of the number of satellites of the central's morphology 
and the redshift is also shown in  Fig. \ref{fig:m_n_com_cen}, where we
included the satellite counts within $R_\mathrm{vir}$ in pale colours in 
the right panels.
The change compared to the number of satellites within $r=300
\mathrm{kpc}$ is only marginal, since the average virial radius of the
central galaxies has about the same size as the considered sphere.

We conclude that counting the satellites of galaxies
without distinguishing between central galaxies and companions,
i.e. large satellites, just reflects the environment they live 
in. The number of satellites within the virial radius of
{\it centrals}, however, remains almost constant with increasing
redshift and therefore traces the underlying dark matter halo.


\begin{figure*}
  \begin{centering}
    \includegraphics[angle=270, width=0.9\textwidth,clip=true]{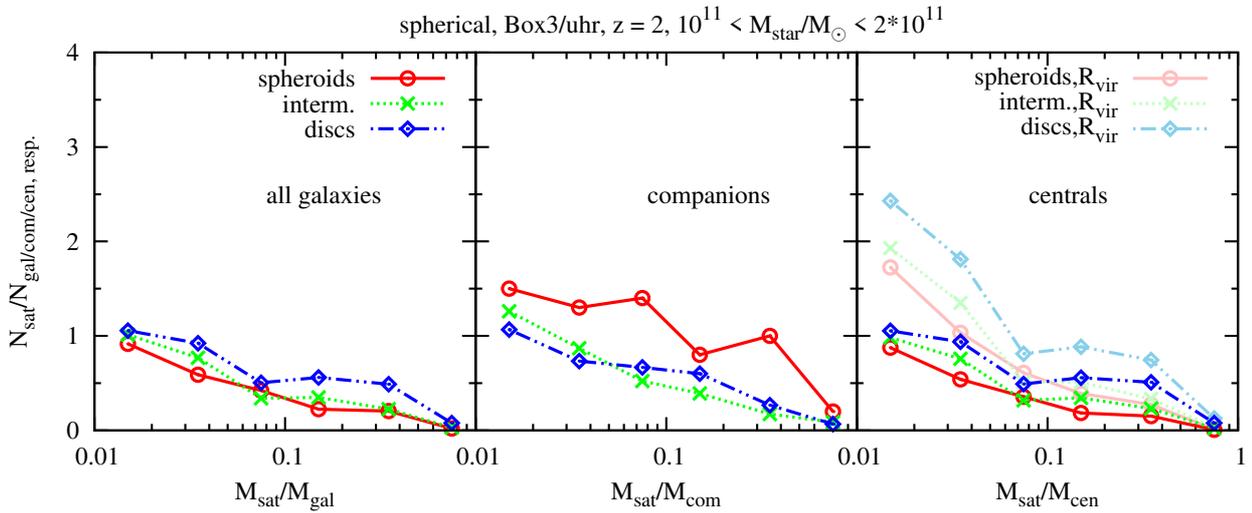}
    \caption{\textit{From left to right:} same as Fig. \ref{fig:m_n_com_cen} but in the larger Box3 with the same resolution at $z=2$.
    There is no split-up of the distribution for all host galaxies (left panel), since the number of companions is very small in this mass range at $z=2$ and thus the distribution for all galaxies is reflecting that of the centrals (right panel). 
    The distribution of the companions exhibits again the split-up (middle panel).}
    \label{fig:m_n_diffr_box3}
  \end{centering}
\end{figure*}

\begin{table}
\caption{The number of all host galaxies with stellar masses higher between $10^{11} M_{\odot}$ and $2 \cdot 10^{11} M_{\odot}$ at $z=2$ in Box3/uhr}
\label{tab:no_comp_cent_box3}
\centering
 \begin{tabular}{l l l l}
 \hline \hline
  galaxy type & $N_\mathrm{spheroid}$ & $N_\mathrm{interm.}$ & $N_\mathrm{disc}$ \\ \hline
  all galaxies & 156 & 209 & 180 \\
  companion & 10 & 23 & 15 \\
  central & 146 & 186 & 165 \\
 \hline  
 \end{tabular}
\end{table}

\subsection{Massive Galaxies at High Redshift}\label{sec:box3}

In order to test the signal for massive galaxies in more detail
by choosing a very narrow mass range of $10^{11} M_{\odot}$ and 
$2 \cdot 10^{11} M_{\odot}$, as done in the observations 
by \cite{Ruiz15}, we now select galaxies from the larger volume simulation 
{\it Box3}. This large volume simulation was so far only evolved down to 
a redshift of $z=2$. This simulation has the same resolution as the simulation 
used before, and we find the same results for the satellite fractions using 
the full mass range. This volume is actually large enough to obtain
several hundreds of galaxies, even within such a narrow mass range, which we can
split into companion and central galaxies and classify them as spheroids, 
intermediates, and discs, as reported in Table \ref{tab:no_comp_cent_box3}.
Note that the underlying AGN feedback in this new, larger volume simulation
has slightly different parameters, which generally leads to a fraction of 
passive galaxies even closer to the observed one, as shown in \cite{Steinborn15}.

In Fig. \ref{fig:m_n_diffr_box3} we plot the mass fraction of the
satellites against the number per host galaxy as in
Fig. \ref{fig:m_n_com_cen} but for the narrow mass range 
of $10^{11} M_{\odot}$ and $2 \cdot 10^{11} M_{\odot}$, where our
simulation resolves the complete mass range of the satellite galaxies. 
We still see the increased number of satellites for elliptical companions,
and no increased number for centrals. As shown before, at this high mass range,
the signal for the centrals even slightly reverses. 
The pale curves again show the satellite number per central within 
the virial radius. From this we conclude that our key result, 
namely that the split-up is caused by the companion galaxies, is not 
biased by a mass selection effect, since we find this split-up in both 
galaxy samples despite the different mass ranges.

\section{Relation between Host and Satellite Properties}\label{sec:SFRs}

Until now we have only considered the number of satellites for centrals
of different morphology. However, at high redshift morphology (traced via 
kinematic properties like the $b$-value) and starforming activity can mean different things, 
as shown before \cite[see also][]{Teklu15}. Additionally, the relation between satellites
and their centrals seems to be more complex.
Therefore, it is important to understand the connection between properties of host and
satellite galaxies in more detail.
In the following section we will take a closer look at the relations between host and satellite galaxies
with respect to their star formation rates, formation redshifts, and dynamical properties.


\begin{figure}
    \begin{centering}
    \includegraphics[angle=270, width=0.45\textwidth,clip=true]{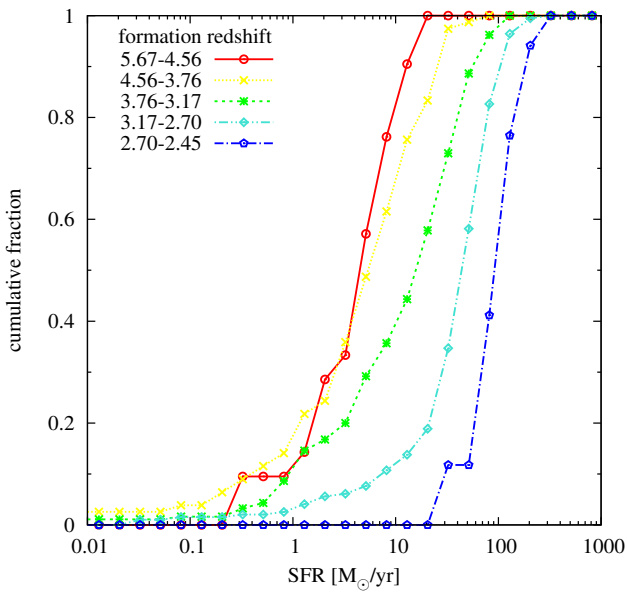}
    \caption{The cumulative fraction of centrals binned according to their SFR in Box3 at $z=2$. 
    The formation redshift of the centrals is colour-coded, where younger centrals are blue and older ones are red. 
    There is a trend that young centrals have a higher SFR than older ones.}
    \label{fig:cen_sfr_age_box3}
  \end{centering}
\end{figure}


\subsection{SFR and Formation Redshift of Satellites and Central Galaxies}

In Fig. \ref{fig:cen_sfr_age_box3} we show, how the current star formation 
rate is reflected in the formation redshift ($z_\mathrm{form}$). To avoid bias
from possible mass-dependence effects, we therefore again analyse the centrals from the
narrow mass range $[10^{11},2 \cdot 10^{11}] M_{\odot}$. 
Here we show the cumulative fraction of centrals according to their 
star formation rate at $z=2$, where the colours represent 
the average redshift, 
where the stars of a given galaxy were formed. As expected, we clearly 
find a smooth, continuous tendency for older (red) centrals to have 
lower star formation rates at current time (e.g. $z=2$), and therefore 
older centrals are more quenched than younger (blue) centrals.
 
In the following we want to verify if and how the $b$-value (i.e., the
morphology), the star formation rate, and the formation redshift of the central galaxies 
correlate with the formation redshift and the star formation of their satellite galaxies
(within the virial radius). 

\begin{figure*}
    \begin{centering}
    \includegraphics[width=0.9\textwidth,clip=true]{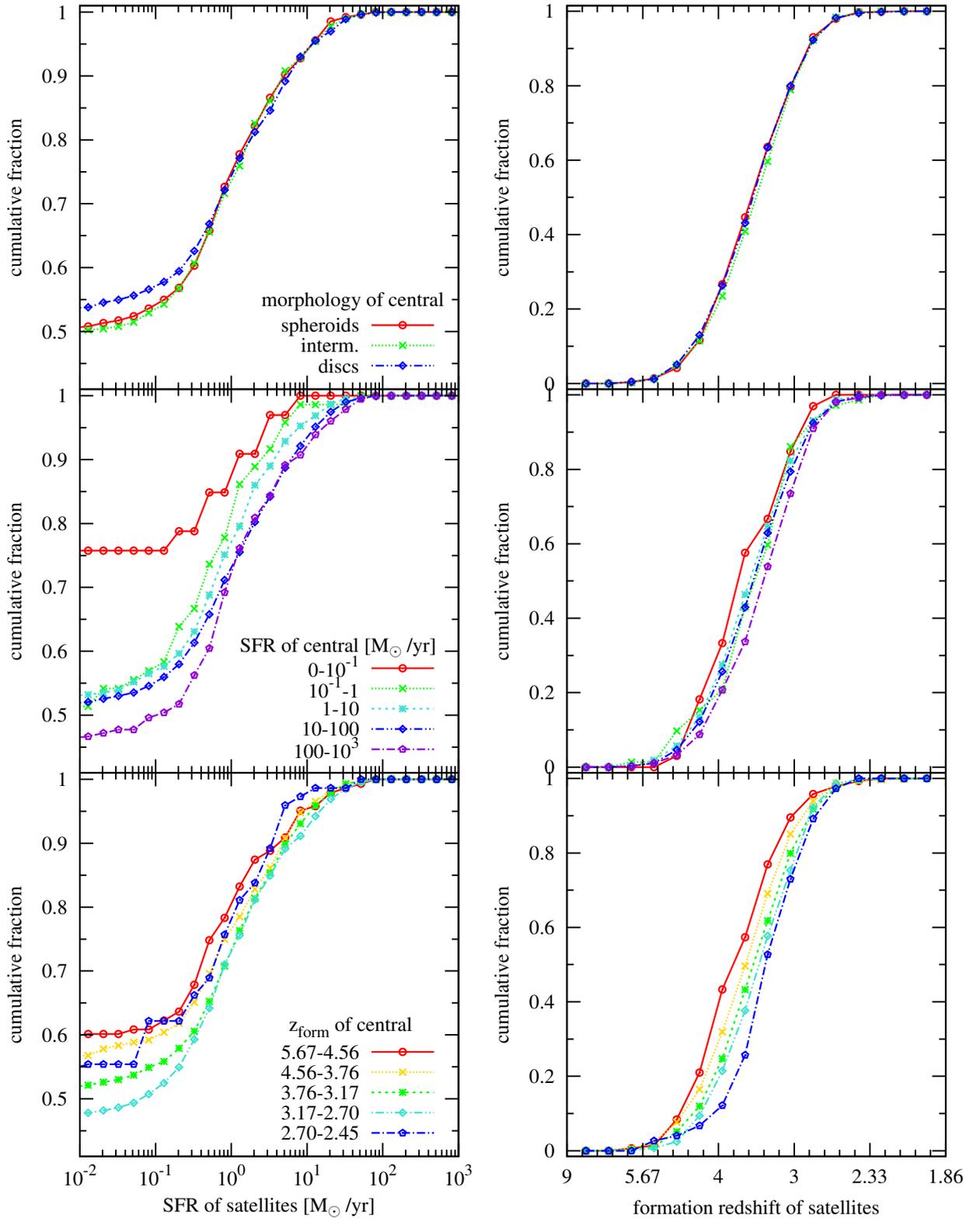}
    \caption{\textit{Left:} The cumulative fraction of satellites within the virial radii of centrals binned according to their SFR. \textit{Right:} The cumulative fraction of satellites according to their formation redshift. 
    In the upper row the colour encodes the $b$-value of the centrals of the satellites, the middle row divides the centrals according to the SFR and the lower panels show the satellite population split according to the formation redshift of their central, for galaxies in Box3 at $z=2$. }
    \label{fig:sat_sfr_age_box3}
  \end{centering}
\end{figure*}

The left column of Fig. \ref{fig:sat_sfr_age_box3} always shows the
cumulative fraction of the star formation of the satellite galaxies, while the 
right column always shows the cumulative fraction of the formation redshift of the 
satellite galaxies. The different colours in the rows reflect different
selection criteria of the underlying central galaxies as follows: 

The upper row shows the sample of host galaxies split into
the different morphological types (e.g. spheroids, intermediates and discs
classified via low, medium and high $b$-values, respectively).
For all host morphologies, the majority of satellite galaxies are already 
quenched and do not show any differences in their formation time.
The curves for the star formation rates for satellites of the disc-like and 
spheroidal centrals are indistinguishable above $0.1M_{\odot}/yr$, and a KS-test 
gives a probability of $5.36 \cdot 10^{-1}$ that the overall distributions (red
and blue curve) are the same. 

In the middle panels we check how the star formation of the centrals is related
to the star formation and the formation redshift of their satellites. The left panel 
shows that centrals with almost no star formation (red curve) are mainly
surrounded by satellites which themselves have no star formation, while
star-forming centrals have a larger fraction of satellites with significant
star formation. 
This suggests that the environment provides the star-forming centrals as
well as their satellites continuously with gas.
Additionally, we see an indication for a weak correlation between the star 
formation of the centrals and the formation redshift of their satellites (right panel). Here, the KS-test gives a
probability of $2.14 \cdot 10^{-2}$ that the two distributions of
non-star-forming (red curve) and highly star-forming (purple curve)
centrals have the same underlying distribution. 

In the lower row we colour-code the distributions according to the formation redshift of the centrals. On the left panel we note a slight correlation
between the star formation of the satellites and the formation redshift of the centrals:
Older central galaxies have a larger fraction of quiescent satellites, younger
centrals have a larger fraction of star-forming satellites, except
for the bin with the youngest centrals (blue curve) which, however, is 
based on low number statistics. On the right panel we find a
correlation between the formation redshift of the satellites and the formation redshift of
their centrals. The older centrals (red curve) have older satellites,
while the younger centrals (blue curve) tend to have younger
satellites, which is confirmed by the KS-test with a probability of
$1.85 \cdot 10^{-6}$. 

We conclude that the morphology of the central galaxy does not relate to
the star formation and the formation redshift of its satellites at $z=2$. 
Nevertheless, we find a significant correlation between the star formation 
and the formation redshift of the centrals themselves with the star formation 
and the formation redshift of their satellites, reflecting the so-called 
conformity \cite[see e.g.][]{Tinker17}.
The fact that there is neither a correlation between the $b$-value of 
the central and the star formation and formation redshift of the satellites nor a correlation 
between the $b$-value and the number of satellites suggests that the 
environment is the main mechanism that controls the global star formation 
properties and thus links the formation redshift of both centrals and satellites. 

\subsection{Evolution of the SFR-$z_{\rm{form}}$ relation}
  
\begin{figure*}
  \begin{centering}
    \includegraphics[angle=270, width=0.95\textwidth,clip=true]{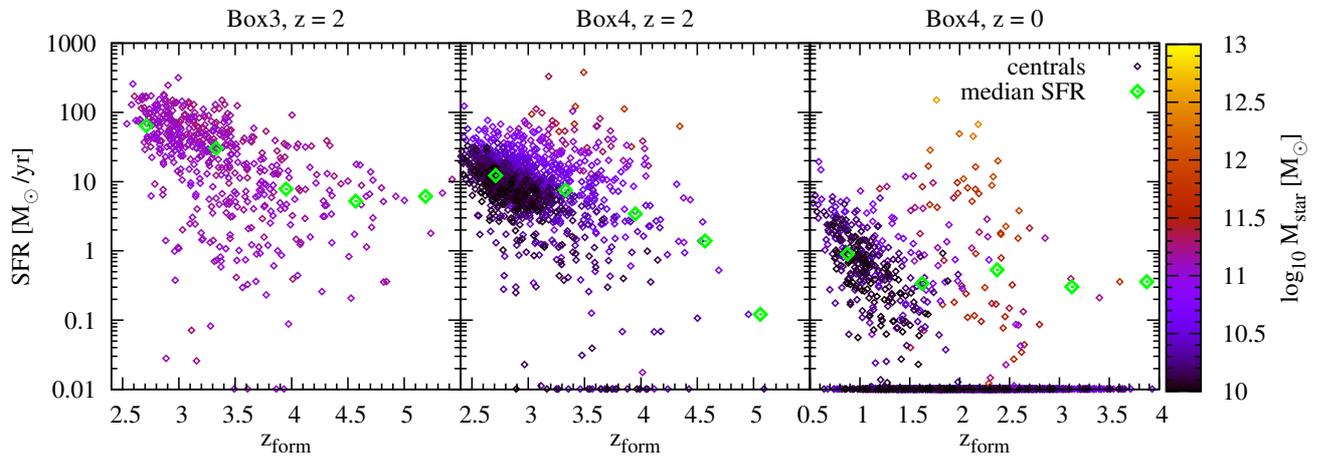}
    \caption{The formation redshift $z_\mathrm{form}$ against the current SFR for centrals with stellar masses between $10^{11}$ and $2 \cdot 10^{11} M_\odot$ in Box3 at $z=2$ (left) and centrals with stellar masses above $10^{10} M_\odot$ in Box4 at $z=2$ (middle) and at $z=0$ (right). }
    \label{fig:cen_sfr_z_m}
  \end{centering}
\end{figure*}

The evolution of the relation between the star formation rate and the formation 
redshift of the central galaxies is shown in Fig \ref{fig:cen_sfr_z_m}.
Here all panels show a scatter plot of the star formation rate
as function of formation redshift, coloured by the stellar mass of the central. 
The left and middle panels present the $z=2$ results for the narrow mass range 
sample from  {\it Box3} (left) and the full sample from the smaller {\it Box4} 
simulation (middle). The right panel shows the according result from {\it Box4} at $z=0$.
The green diamonds are the averaged star formation rate, binned for different 
formation redshifts. Centrals with star formation rates
lower than $0.01M_{\odot}/yr$ are plotted on the lowest displayed
value of the $y$-axis.  

We find a clear trend of the median star formation rate with the formation
redshift at $z=2$, in agreement with our previous 
finding that old centrals tend to have a lower star formation rate, 
while young centrals on average have a higher star formation rate. 
A part of this relation is expected, as more massive galaxies are expected to have
higher star formation rates and also form earlier. However, this relation
holds even at fixed stellar mass, as shown by the narrow mass 
range sample (left panel in Fig. \ref{fig:cen_sfr_z_m}).
This means that galaxies with already quenched star formation 
usually have formed earlier, in line with previous studies by e.g. \citet{Feldmann16}.  
At $z=2$ we find a median formation redshift for
star-forming galaxies to be 3.36, which is in line with recent 
observations by \citet{Thomas16}, who evaluate a
median formation redshift of $\sim 3.22$ for their galaxy sample
observed at $2<z<6.5$ when using a formation redshift based on mean stellar ages similar to ours.

At redshift $z=0$ (right panel), we still find this correlation between
the formation redshift $z_\mathrm{form}$ and the star formation rate for low-mass galaxies, while high-mass galaxies show no clear trend anymore.
These massive galaxies live in very dense environments like galaxy groups 
and clusters in which star formation can continue, fed by cooling from the 
halo, even for systems with early formation times. Such massive galaxies at $z=0$
all have formed at early times, typically with a formation redshift between 
$z=1.5$ and $z=2.5$.

\begin{table}
\caption{The number of centrals and companions with $M_\mathrm{star} > 10^{10} M_{\odot}$ divided into quiescent and star-forming at four different redshifts}
\label{tab:no_cent_qu_sf}
\centering
 \begin{tabular}{l | l l | l l}
 \hline \hline
   & centrals & & companions &   \\
 \hline \hline
  redshift & $N_\mathrm{quiescent}$ & $N_\mathrm{SF}$ &  $N_\mathrm{quiescent}$ & $N_\mathrm{SF}$\\ \hline
  0 & 1069 & 250 & 694 & 99 \\
  0.5 & 1064  & 380 & 633  & 148     \\
  1 & 908  & 636 & 466  & 200    \\
  2 & 107  & 1198 & 73  & 242 \\
  \hline 
 \end{tabular}
\end{table}


\begin{figure}
  \begin{centering}
    \includegraphics[width=0.49\textwidth,clip=true]{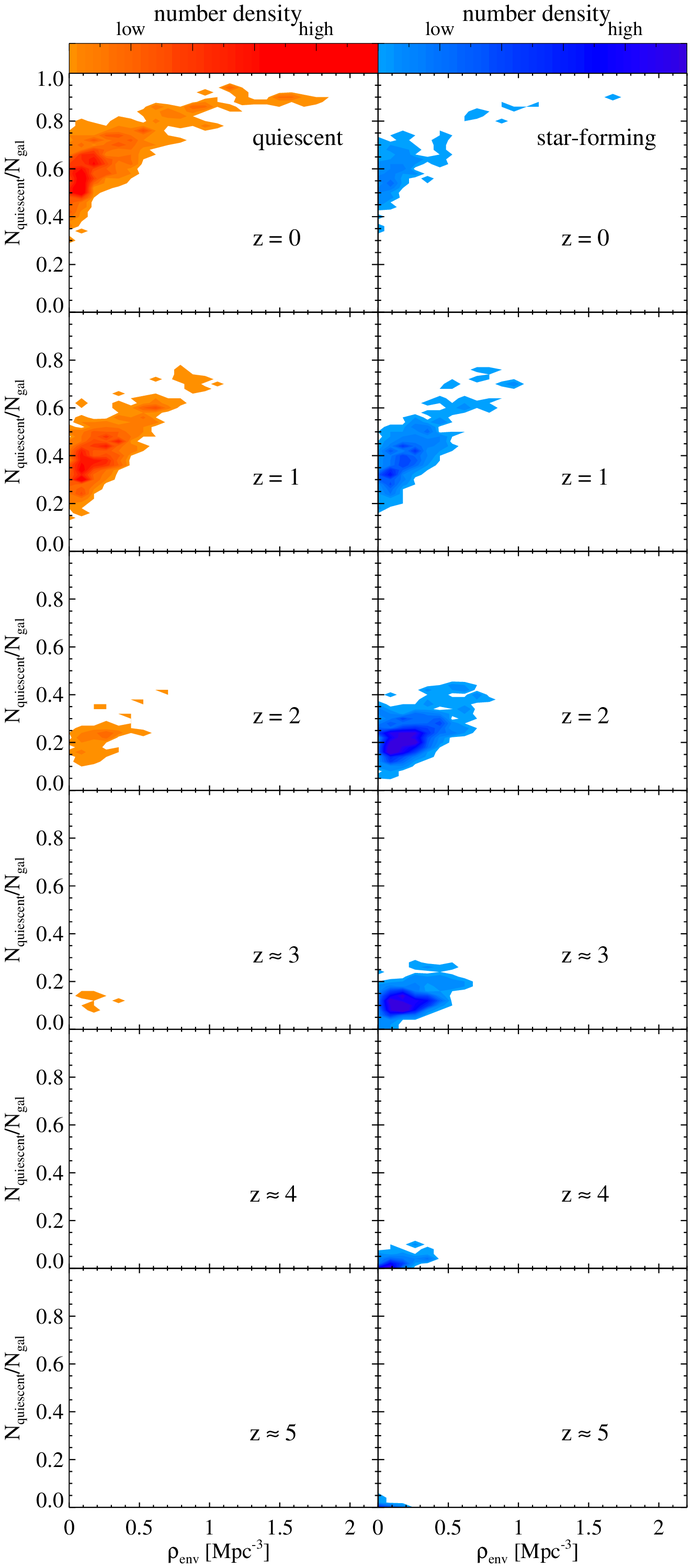}
    \caption{The fraction of quiescent galaxies more massive than $10^8 M_\odot$ within a sphere of radius 5 Mpc (co-moving) around quiescent (left panels)  and star-forming (right panels) centrals with stellar masses above $10^{10} M_\odot$ in Box4 at different redshifts. There is a clear build-up of the density-morphology-relation starting at around $z=2$; while at high redshift the fraction of quiescent galaxies does not depend on the number of neighbouring galaxies, i.e. the environment, with decreasing redshift the environment becomes more and more important for the quiescent fraction. The distribution looks very similar for quiescent and star-forming galaxies.}
    \label{fig:cen_ngal_fqu}
  \end{centering}
\end{figure}

\subsection{The Relation between the Environment and the Star Formation Rate}

To investigate how the environment is affecting the star formation rate of galaxies, 
we characterize the environment of galaxies by their neighbour counts. Therefore, we count 
all galaxies with stellar masses above $10^8 M_\odot$ within a sphere of
radius $r=5~{\rm{Mpc}}$ around our centrals. We define the environmental density as
\begin{equation}
  \rho_\mathrm{env}=\frac{N_\mathrm{gal}(5Mpc)}{4/3 \cdot \pi \cdot
    r^3}.
\end{equation}
Here, following \citet{Treu03}, the radius was chosen to include all potential 
cluster members. Accordingly, using the distance to the 10th neighbour results 
in a similar median value \citep[see][]{Cappellari11a}.

Fig. \ref{fig:cen_ngal_fqu} shows the build-up of the density-morphology-relation 
\citep[e.g.][]{Dressler80} with cosmic time. The different panels show the
dependence of the density-morphology-relation on the star formation properties
of the central galaxies: The left panels show the quiescent centrals, while the
right panels show the star-forming centrals. Table \ref{tab:no_cent_qu_sf}
lists the sample sizes at the different redshifts.

We do not find evidence for a density-morphology-relation
at high redshift ($z > 2$), for neither of the two galaxy types.
Generally, we see a clear evolution with decreasing redshift, and the
build-up of the density-morphology-relation starts at around $z=2$.
At this redshift the trend appears 
that centrals that live in over-dense regions have a high
fraction of quiescent neighbours, while galaxies in lower density
environments have on average a lower fraction of quiescent neighbours.
This trend is in good agreement with observations by
\citet{Darvish16}, who also find that at higher redshift the fraction
of quiescent galaxies does not depend on the environment, while it
does at low redshift. Interestingly, it 
does not make a difference if the central is quiescent or star-forming
itself: There are for example quiescent centrals in low-density
environments that have a very low fraction of quiescent galaxies
around them and vice versa, star-forming centrals in high-density
environments with a high fraction of quiescent galaxies.

\begin{figure}
  \begin{centering}
    \includegraphics[width=0.49\textwidth,clip=true]{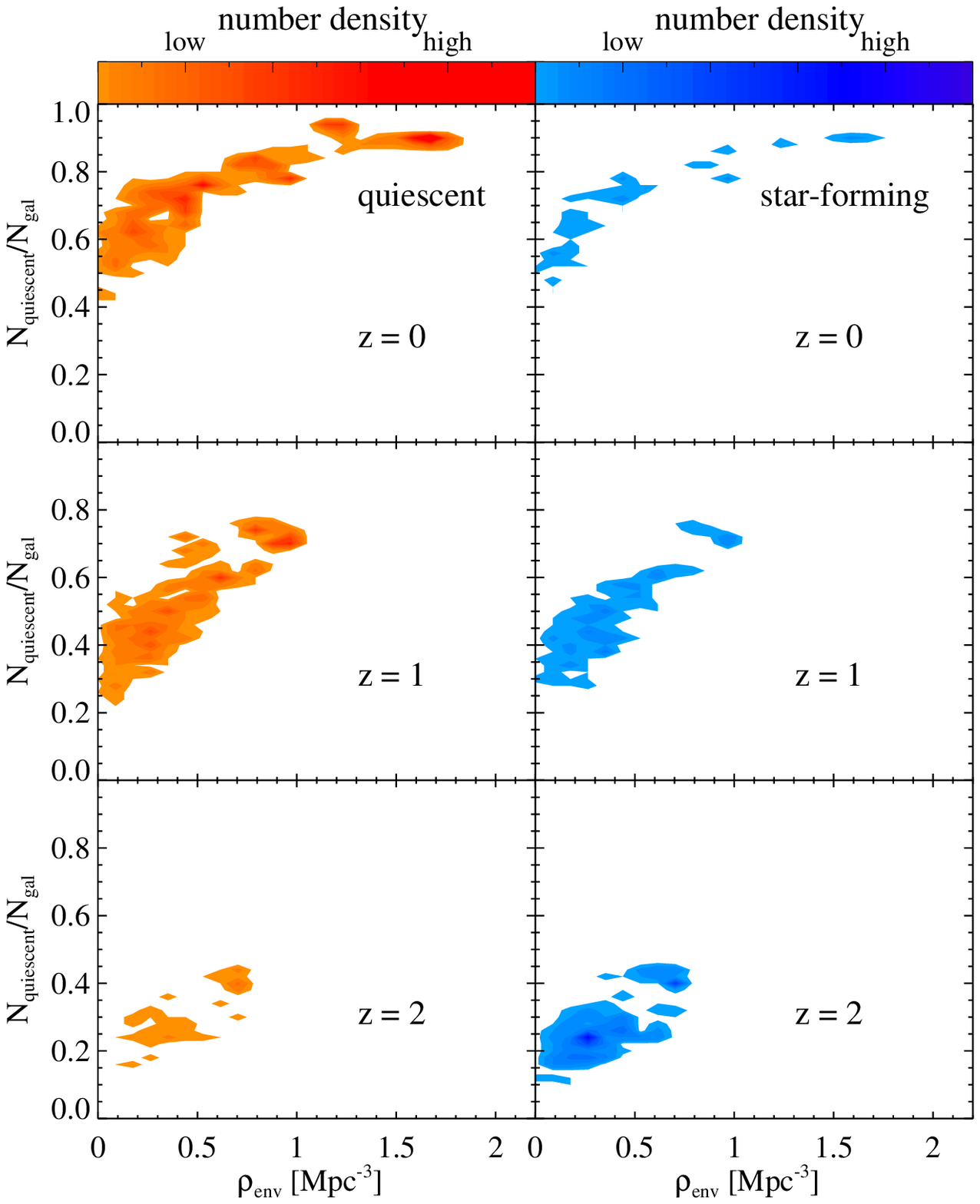}
    \caption{The same as in Fig. \ref{fig:cen_ngal_fqu}, but for companions. As seen for the centrals, there is a built-up of the density-morphology-relation, where there is no difference between quiescent and star-forming companions. }
    \label{fig:comp_ngal_fqu}
  \end{centering}
\end{figure}

Several studies (e.g., \citet{Peng10}, \citet{Lee15} and \citet{Hirschmann16}) 
find that this environmental quenching is important especially for 
low-mass satellite galaxies. 
For completeness, we show in Fig. \ref{fig:dmr_phys} of the Appendix the density-morphology-relation for centrals in physical units. 

In Fig. \ref{fig:comp_ngal_fqu} we show the density-morphology-relation
for the companion galaxies for redshifts $z=0$ to $z=2$, as there are too
few companions at higher redshifts.  
The shape of the density-morphology-relation for the companion galaxies 
is the same as for the central galaxies (see Fig. \ref{fig:cen_ngal_fqu}).
Similarly, it does not make a difference for the fraction of quiescent 
neighbouring galaxies if the companion galaxy itself is quiescent or star-forming.   
While the number density of the quiescent central galaxies peaks in the low-density 
environments, the number density of the quiescent companions is more shifted towards 
denser environment. In contrast, the distribution of the number densities in the 
density-morphology-relation of star-forming centrals and satellites does not differ. 
This clearly demonstrates that environmental quenching is more important for satellite
galaxies than for centrals. 
Since our satellite galaxies on average are low-mass galaxies this supports earlier
studies by \citet{Hirschmann16}, who concluded that the fraction of quiescent galaxies
is mainly determined by internal processes, while the environment becomes important at 
lower masses.


\section{Discussion and Conclusions}\label{sec:conclusion}

We use galaxies extracted from the high resolution cosmological
simulation suite Magneticum to investigate the relation between the
dark matter halo mass and the internal properties of galaxies  as well
as the environmental imprint on the star formation properties.
Especially, we want to test whether the morphological type of a galaxy depends on the dark matter halo mass, 
as recently concluded from the
observed differences in the abundance of satellites around galaxies of
different morphological types by \citet{Ruiz15}.  We select all
galaxies with stellar masses above $10^{10} M_{\odot}$  and use a
dynamical parameter, the so-called $b$-value \citep[see][]{Teklu15},
as indicator of morphological type. We find that:

\begin{itemize}

\item When applying the observed strategy of counting satellite
  galaxies around all massive galaxies we clearly reproduce the
  observed signal of more satellites around spheroidal galaxies
  compared to disc galaxies for both the total numbers for different
  mass ratios as well as for the radial abundance profile.

\item This signal does not depend on (and therefore is not produced
  by) the way of defining the volume in which the satellites are
  counted. We find that the signal even holds if we use spherical
  volumes around the galaxies instead of a cylinder defined by a
  redshift range as observers have to rely on.

\item However, when splitting the sample of massive galaxies into
  galaxies at the centre of their dark matter halo (i.e. centrals) and
  companion galaxies (which are only large satellites of other massive
  galaxies within a common dark matter potential), we find that the
  signal is exclusively caused by the companion galaxies. If only the
  central galaxies are considered, the signal completely disappears.

\item This result is found to hold true even at higher redshifts up to
  $z=2$. However, when considering all galaxies (centrals and companions),
  the difference in the satellite number between
  spheroids and discs becomes smaller with increasing redshift due
  to the lower number of companions at higher redshift.

\item This is also supported by the fact that the baryon conversion
  efficiency of our galaxies does not show any significant trend
  with the inferred morphological type. 
\end{itemize}

We therefore conclude that the observed differences in the abundance
of satellites around massive galaxies of different morphological types
is only driven by the different environments, in which spheroid and disc
galaxies are typically living, and that they are not related to
differences in the baryon conversion efficiency of galaxies with
different morphological types. This is supported by observational
results from \citet{Guo15} that isolated primaries in a filamentary
environment have more satellites than those outside of filaments.

We investigate in more detail the relation between the quenching of
star formation and our morphological classification based on the
$b$-value for galaxies at high redshift (i.e. $z=2$). Selecting a
large number of galaxies ($\approx 500$) within a very narrow mass
range (i.e. $10^{11} M_\odot < M_\mathrm{star} < 2\cdot10^{11}M_\odot$), 
allows us to exclude any mass dependencies. We find that:

\begin{itemize}

\item In contrast to galaxies at $z=0$, the galaxies which are 
  already quenched at $z=2$ do not show a relation to their 
  morphological type.

\item Quenched galaxies at $z=2$ have on average formed
  earlier than their starforming counterparts, as suggested in previous studies
  \citep[e.g.][]{Feldmann16}. 

\item Satellites of quenched galaxies at $z=2$ have typically formed earlier 
  and show on average less star formation activity compared to satellites of 
  star-forming galaxies, as suggested in earlier studies \citep[e.g.][]{Feldmann16}.
\end{itemize}

Our simulations furthermore show that at redshifts of $z=2$, in
contrast to present time, the dynamical classification of galaxies 
based on morphology results in a different selection than the classification 
based on star formation. Our results are broadly in agreement with the
picture that was pointed out in previous studies \citep[e.g.][]{Peng10,Hirschmann16,Huertas16} that the environment is
effective in quenching star formation of lower mass galaxies, while at
higher masses and redshifts $z > 1$ other (internal) effects are the 
dominant drivers.

Additionally, we study the connection between the environment and the star 
formation rate evaluating the environment density $\rho_\mathrm{env}$. From the
evolution of the density-morphology-relation between $z \approx 5$ and $z=0$
we conclude that:

\begin{itemize} 
\item At high redshifts ($z\geq3$) there is no signature of a density-morphology-relation for central galaxies. The built-up of the density-morphology-relation 
  starts at around $z\approx2$.
\item The density-morphology-relation for quiescent and starforming centrals 
  is similar, indicating a negligible influence of the environment on the
  star-forming properties of the central galaxies. 
\item The shape of the density-morphology-relation for the companion galaxies 
  is the same as for the central galaxies.
\item The quiescent fraction of companion galaxies is comparable to that of 
  the central galaxies.
\item The number density of the quiescent central galaxies peaks in the low-density 
  environments, while the number density of the quiescent companions peaks at 
  denser environments. 
\item The distribution of the number densities in the density-morphology-relation 
  of star-forming centrals and satellites does not differ.
\end{itemize} 

We conclude that environmental quenching is more important for satellite
galaxies than for centrals.


\section*{Acknowledgements}

We thank Lisa Steinborn, Tadziu Hoffmann and Felix Schulze for useful discussions. 
AFT and KD are supported by the DFG Research Unit 1254 and the DFG Transregio TR33. 
AB is supported by the DFG Priority Programme 1573.
This research is supported by the DFG Cluster of Excellence `Origin and Structure of the Universe'. 
Computations have been performed at the `Leibniz-Rechenzentrum' with CPU time assigned to the Project `pr86re'.
We are especially grateful for the support by M. Petkova through the Computational Center for Particle and Astrophysics (C2PAP).

\bibliographystyle{mnras}
\bibliography{bibliography}

\appendix

\section{The Radial Distribution of Satellites}

\begin{figure}
  \begin{centering}
    \includegraphics[angle=270,width=0.49\textwidth,clip=true]{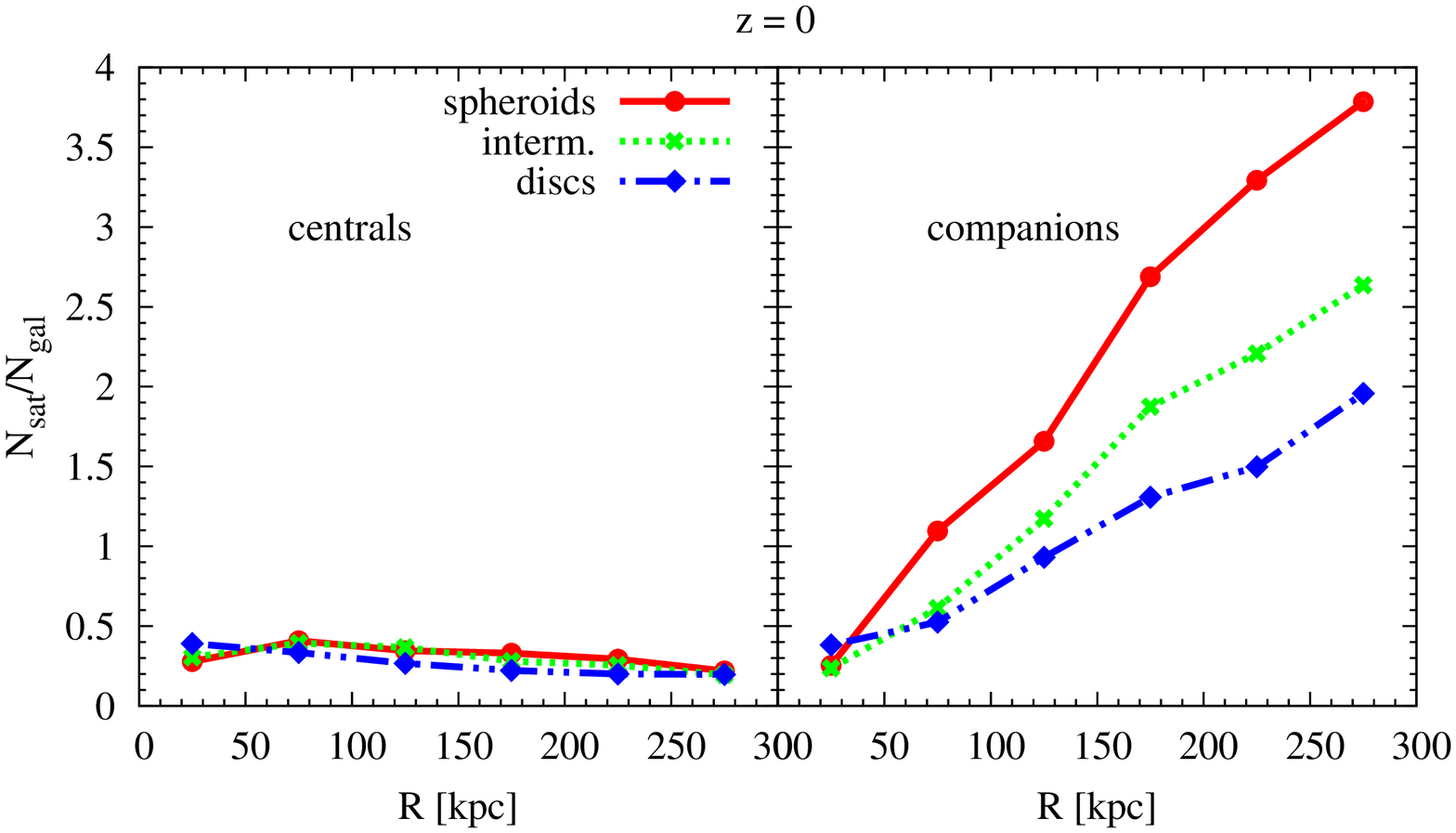}
    \caption{The differential radial distribution of satellites per
      host galaxy at redshift $z=0$ for centrals (left) and companions
      (right).}
    \label{fig:r_n_cen_com}
  \end{centering}
\end{figure}

Previously we showed the apparent difference in the radial profiles of 
satellite abundance around spheroidal and disc-like galaxies, where spheroidal 
galaxies have more satellites than disc-like host galaxies (Fig. \ref{fig:n_r_z}). 
We demonstrated that this split-up is caused by 
the companion galaxy population (see middle and right panels Fig. \ref{fig:m_n_com_cen}).
In Fig. \ref{fig:r_n_cen_com} we demonstrate that this difference in the 
radial distribution of satellites around the host galaxies also originates
purely from the companions (right panel), while the signal completely
disappears for the central galaxies (left panel), as expected.

\section{The World in Physical Units}

Since it is common to use physical units in observations, we show some
of our results in physical units, where we take the expansion of the
Universe into account. The difference to the co-moving units, which
capture the real growth of a structure, becomes more and more
important with increasing redshift.

\subsection{Satellite Fraction in Physical Units}

\begin{figure}
  \begin{centering}
    \includegraphics[angle=270,width=0.49\textwidth,clip=true]{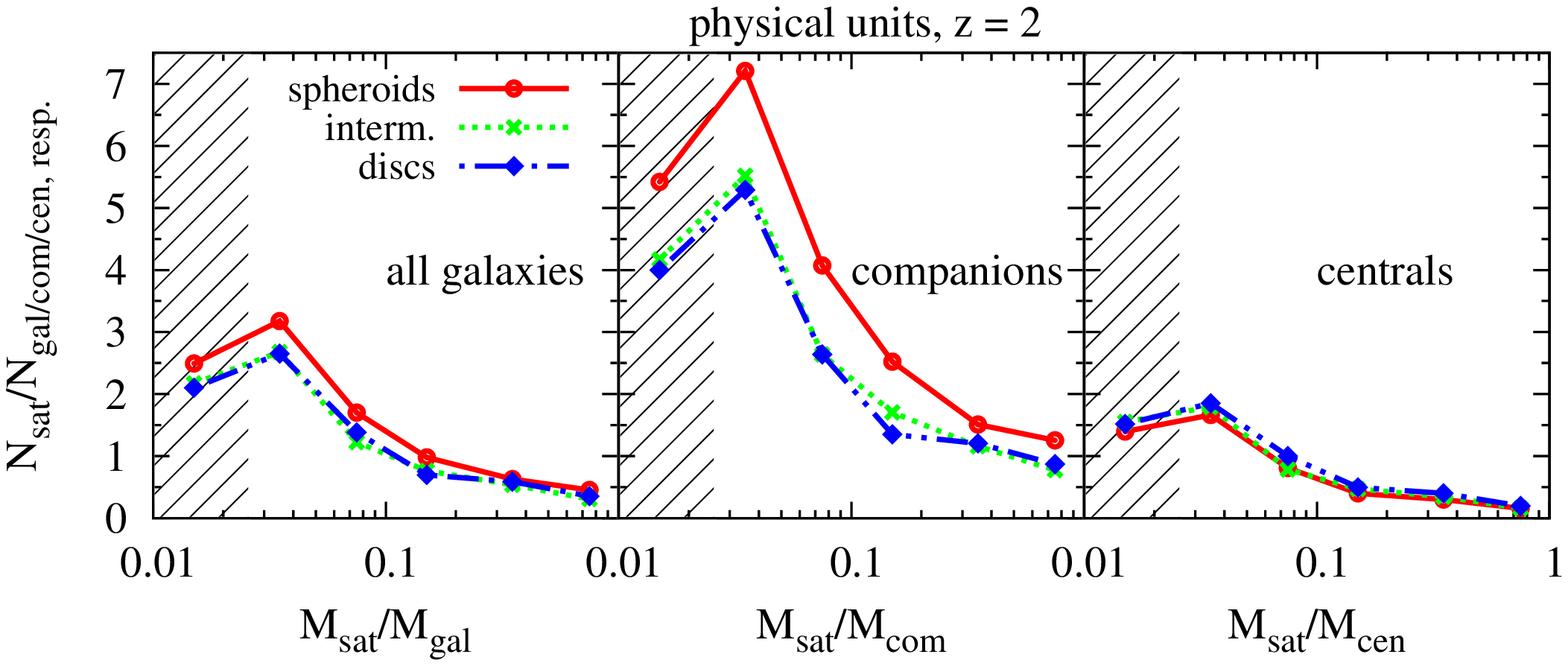}
    \caption{The differential number of satellites per host galaxy at
      redshift $z=2$ for all host galaxies (left), companions (middle)
      and central galaxies (right) within a sphere of 300 kpc in
      physical units, binned according to their mass ratio.}
    \label{fig:m_n_phys}
  \end{centering}
\end{figure}

In Fig. \ref{fig:m_n_com_cen} we showed the abundance of satellites
per host galaxy binned by the mass ratio with respect to their host,
for all potential host galaxies (left panel) and split into companion
(middle panel) and central (right panel). Fig. \ref{fig:m_n_phys}
shows the same for redshift $z=2$ in physical units. Compared to the
bottom panels in Fig. \ref{fig:m_n_com_cen}, the number of satellite
galaxies is higher. It is also higher than at $z=0$, which is not
surprising when comparing volume-limited samples at different
redshifts (see also discussion in \citet{Conroy07}). 
This demonstrates that careful choice of distance scale is important, 
especially at high redshift.

\subsection{Density-Morphology-Relation in Physical Units}

\begin{figure}
  \begin{centering}
    \includegraphics[width=0.49\textwidth,clip=true]{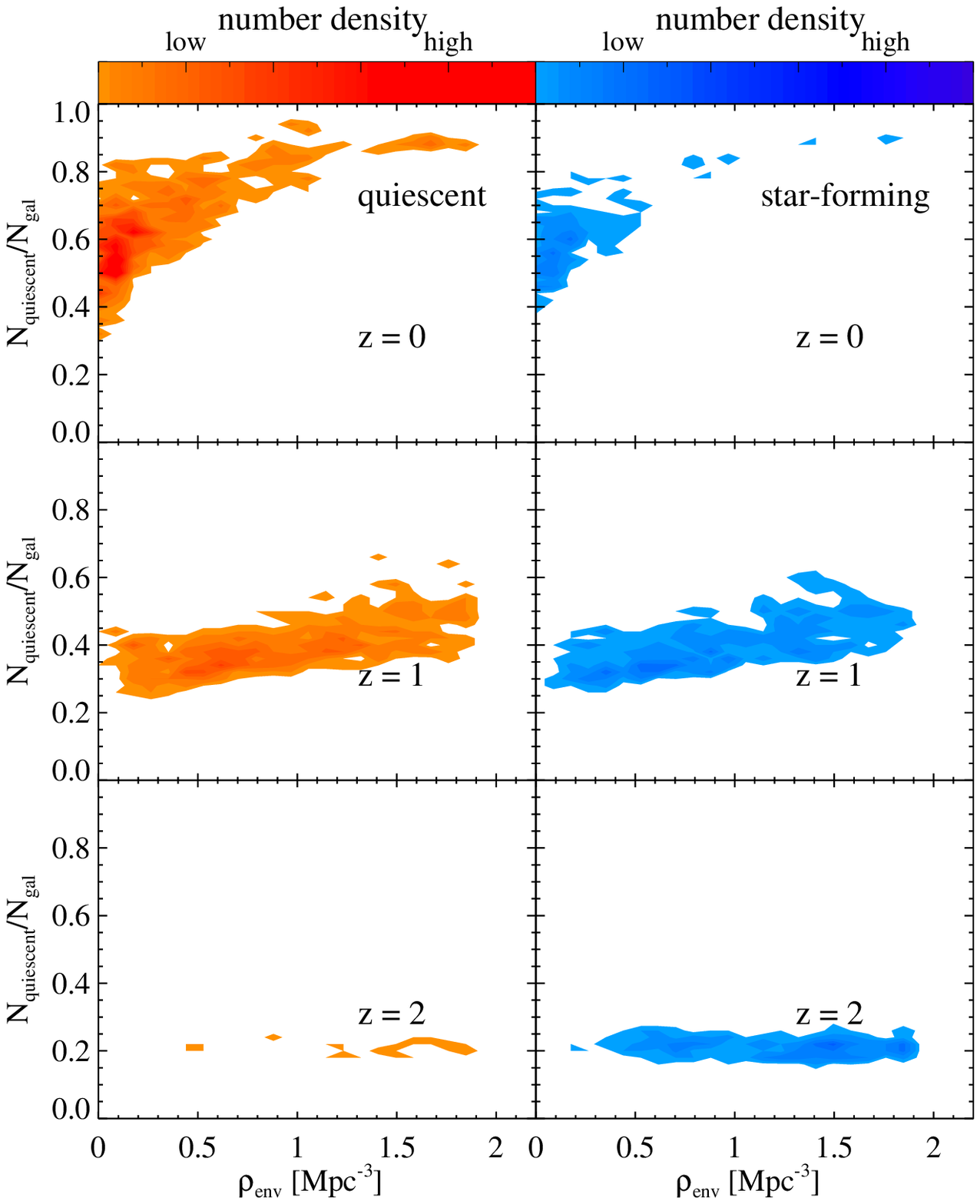}
    \caption{The fraction of quiescent galaxies more massive than
      $10^8 M_\odot$ within a sphere of radius 5 Mpc (physical) around
      quiescent (left panels) and star-forming (right panels) centrals
      with stellar masses above $10^{10} M_\odot$ in Box4 at different
      redshifts.}
    \label{fig:dmr_phys}
  \end{centering}
\end{figure}

The influence of the choice of distance scale becomes especially
prominent for the density-morphology-relation.
Comparing Fig. \ref{fig:cen_ngal_fqu}, which illustrates the
density-morphology-relation for centrals in co-moving units, with
Fig. \ref{fig:dmr_phys}, which is the same relation in physical units, 
the shape and its evolution look strikingly different.
When taking a sphere in physical units, the induced change in the distance
scale masks the presence of the density morphology relation at $z\geq1$,
resulting in a later and steeper increase of the density-morphology-relation. 
This is due to
the fact that in physical units the search radius becomes larger
compared to the central galaxies with increasing redshift and we thus
go farther outside of the ``local'' environment to the ``global''
environment. 
In the ``local'' environment we find a higher number of
quiescent neighbouring galaxies than in the ``global''
environment, emphasizing the importance of the ``local'' 
environment for the satellite quenching.

\bsp
\label{lastpage}
\end{document}